\documentclass[journal,draftcls,onecolumn,12pt]{IEEEtran}

\usepackage{graphics} 
\graphicspath{{./picture/}}
\usepackage{float}
\usepackage{tikz}
\usepackage{xcolor} 
\usepackage{lipsum} 
\usepackage{epsfig} 
\usepackage{amsmath} 
\usepackage{amssymb}  
\usepackage{bm}
\usepackage{algorithm}  

\usepackage{algorithmicx}
\usepackage{algpseudocode}
\usepackage{graphicx} 
\usepackage{epstopdf}
\usepackage[short]{optidef}
\usepackage{hyperref}
\hypersetup{linkbordercolor=pink}
\newtheorem{prop}{Proposition}

\newtheorem{remark}{Remark}

\usepackage[subtle]{savetrees}

\begin{document}

\title{
Waveform and Beamforming Design for Intelligent Reflecting Surface Aided Wireless Power Transfer: Single-User and  Multi-User  Solutions
}

\author{

\IEEEauthorblockN{Zhenyuan Feng,  \IEEEmembership{ Member, IEEE}, Bruno Clerckx, \IEEEmembership{Senior Member, IEEE} and Yang Zhao,  \IEEEmembership{ Member, IEEE}}
 \\\vspace{-4mm}
\thanks{The authors are with the Department of Electrical and Electronic Engineering,
Imperial College London, London SW7 2AZ, U.K. (e-mail: {z.feng19,
b.clerckx, yang.zhao18}@imperial.ac.uk).}
}

\maketitle

\begin{abstract}
In this paper, we study the waveform and passive beamforming design for intelligent reflecting surface (IRS)-aided wireless power transfer (WPT). Generalized   multi-user and low complexity single-user algorithms are derived based on alternating optimization (AO) framework to maximize the weighted sum output DC current, subject to the transmit power constraints and passive beamforming  modulus constraints. The input signal waveform and IRS passive beamforming phase shifts  are jointly designed as a function of users' individual frequency-selective channel state information (CSI).  The energy harvester nonlinearity is explored and   two IRS deployment schemes, namely frequency selective  IRS (FS-IRS) and  frequency flat  IRS (FF-IRS), are modeled and analyzed. This paper highlights the fact that IRS can  provide an extra passive beamforming gain on output DC power over conventional WPT designs and significantly influence the waveform design by leveraging the benefit of passive beamforming, frequency diversity and energy harvester nonlinearity. Even though FF-IRS exhibits lower output DC current than the ideal FS-IRS,  it still achieves substantially increased DC power over conventional WPT designs.
Performance evaluations confirm the significant benefits of a joint waveform and passive beamforming design accounting for the energy harvester nonlinearity to boost the performance of single-user and multi-user WPT system.

\end{abstract}


\IEEEpeerreviewmaketitle
\section{Introduction}\label{Section_introduction}
\par
In the last decade, far field wireless power transfer (WPT) has drawn a great attention because of its potential  to energize
  billions of  future autonomous low-power devices. It is deemed as the enabler of the 1G of mobile power networks \cite{clerckx20181GMobile}, and numerous applications such as simultaneous wireless information and power transfer (SWIPT), wirelessly powered communication network (WPCNs), Internet of Things (IoT)
and wireless powered backscatter communication (WPBC) \cite{clerckx2018fundamentals}.
The major challenge of far field WPT is to maximize the output DC power or equivalently
the end-to-end power transmission efficiency (EPTE)  without increasing the transmit power, and for distances over a few  to hundreds of meters. The traditional solution to resolve this challenge has focused on the design of efficient energy harvesters, so-called rectennas, with high RF-to-DC conversion efficiency (CE)
\cite{valenta2014harvesting}\cite{shen2017dual}\cite{shen2018multi}. More recently, in addition to rectenna design, a new complementary research direction on signal designs for WPT has gradually emerged  \cite{7867826}. In the past few years, four different signal design strategies have been proposed and optimized.

\par A \textit{first} strategy is the design of energy \textit{waveforms} to utilize the nonlinearity characteristic of the rectenna and boost jointly the RF-to-RF CE $e_{\text{rf-rf}}$ and the RF-to-DC CE $e_{\text{rf-dc}}$ \cite{clerckx2016waveform}. Such design originates from the fact that 
the output of the energy harvester (EH) is a nonlinear function of the rectenna input signal. Hence, the transmit waveform design has a significant influence on the EPTE. It not only affects the RF-to-RF CE and the signal intensity at the input of rectenna, but also the RF-to-DC CE of the rectifier. A stepping stone in such design was made in \cite{clerckx2016waveform} where a systematic methodology was derived to design and optimize waveforms for WPT. Different from prior non-optimized waveform designs \cite{collado2014optimal}, the optimal waveform design in \cite{clerckx2016waveform} was adapted to the frequency selective channel (with frequency flat channel being a special case) and was rooted in the tradeoff between maximizing $e_{\text{rf-dc}}$ and $e_{\text{rf-rf}}$. Such optimized waveform was shown to provide significant benefits over conventional continuous-wave signal and non-optimized waveforms by leveraging the channel frequency diversity gain and a gain originating from the rectifier nonlinearity. Since the optimal signal design is the solution of computationally involved problems, low complexity methods have been proposed in \cite{clerckx2017low,huang2017large}, and have been experimentally validated in \cite{KimPrototype}. Furthermore, since the channel state information at the transmitter (CSIT) is needed to design the optimal waveform, waveform designs based on limited feedback have been proposed in \cite{huang2017waveform}, and experimentally demonstrated in \cite{shen2021closed}.
\par A \textit{second} strategy is the design of multi  antenna \textit{beamforming} to boost the input power of EH and enhance the RF-to-RF CE $e_{\text{rf-rf}}$. Due to the hardware limitation of EH, acquisition of CSIT is challenging. An efficient method was proposed in \cite{xu2014energy} to overcome this challenge by designing beamforming with limited feedback.
Other directive or enegy focusing solutions, such as time-reversal techniques \cite{ku2017joint} and time-modulated arrays \cite{masotti2016time}, have also been used in WPT system to exploit the real-time antenna beamforming strategies. Beamforming is not limited to the transmitter and can also be used at the receiver subject to a proper design of the combiner scheme \cite{ShenCombining}. Waveform and beamforming can also be combined into a joint waveform and beamforming design \cite{clerckx2016waveform,huang2017large,shen2020joint} to further boost the output DC power and the range of WPT, as demonstrated experimentally in \cite{kim2020range}. Joint waveform and beamforming enables to simultaneously harvest three different gains, namely a beamforming gain, a frequency diversity gain and a gain related to the rectifier nonlinearity, and therefore offers additional opportunities over spatial domain processing/beamforming-only.
\par A \textit{third} strategy is to design \textit{modulation} for WPT. Different from \cite{clerckx2016waveform} where energy waveforms consist of deterministic unmodulated multi-carrier signals, modulation induces random fluctuation of the input rectifier signal to increase $e_{\text{rf-dc}}$. For example, real Gaussian modulation outperforms circularly symmetric complex Gaussian (CSCG) modulation despite the same average input power to the rectifier \cite{varasteh2017wireless}. Similarly, \cite{varasteh2020capacity} has observed that an on-off keying based modulation strategy outperforms real Gaussian modulation in terms of output DC power.
\par A \textit{fourth} strategy is to apply phase sweeping \textit{transmit diversity} in multi-antenna settings \cite{clerckx2018beneficial}. Compared to the beamforming strategy, transmit diversity does not rely on CSIT but still increase $e_{\text{rf-dc}}$ and therefore output DC power.

\par In parallel to the active research area on signal design strategies for WPT, a new paradigm
named \textit{intelligent reflecting surfaces} (IRS) has recently attracted the attention of the wireless research community to further
boost the efficiency of wireless communications \cite{tan2016increasing,hu2018beyond,wu2018intelligent,lee2017retrodirective,shen2020modeling}. Compared with the active beamforming of (massive) multi-antenna systems,
IRS benefits from the presence of passive beamforming reflection elements (REs) integrated into the propagation environment, which are optimized to adapt the wireless channel so as to be favorable to communication receivers. The key advantages of IRS are the tunability of the phase shifters at each RE (so-called passive beamforming phases), the flexibility
of the deployment on arbitrary shaped surfaces and the sustainability due to low cost and power consumption. By collaboratively adjusting the passive beamforming phases, the incident signal is reflected at each RE such that the reflected signals are constructively accumulated at the target receiver to increase the receive signal power, therefore enabling a passive beamforming gain. Nevertheless, in contrast to active antenna arrays where the amplitudes and phases can be adjusted freely at each antenna and at each frequency, REs may be subject to less flexibility due to the passive nature of the IRS and the hardware constraints. Specifically, the  reflection amplitude of the REs is no greater than 1 \footnote{fixed to unity in this paper for maximization of output DC power.}. Moreover the IRS is commonly assumed be consistent with frequency-flat in the sense that passive beamforming phases are kept constant across frequencies, 
(so-called frequency-flat IRS in the sequel). Under those constraints, IRS-assisted wireless communications have focused on single carrier, frequency-flat channel and frequency-flat IRS by jointly designing the active transmit beamforming at the transmitter and passive reflection beamforming at the IRS for single-user \cite{yu2019miso} and multi-user scenarios \cite{wu2019intelligent,Guo2019,pan2020multicell}. Multi carrier OFDM waveform designs were also studied in some papers using frequency-flat IRS \cite{zheng2019intelligent,yang2019irs}. 
\par Aside communications, IRS brings some natural benefits to wireless power since IRS can help increasing the power level at the input of the rectenna. Existing IRS-aided WPT/SWIPT works focus on single tone, frequency-flat IRS, frequency-flat channel and jointly optimize transmit beamforming and passive IRS beamforming phases \cite{pan2020intelligent,wu2019weighted,tang2019joint,wu2019joint}. 
\par The limitation of the existing IRS-aided WPT literature is that the focus is on the spatial domain beamforming-only and that the rectenna nonlinearity has been ignored. Ignoring the rectenna nonlinearity is equivalent to assuming that $e_{\text{rf-dc}}$ is constant and independent of the rectenna input signal power and shape \cite{clerckx20181GMobile,7867826}. Unfortunately, such assumption is well documented to be inaccurate and to lead to inefficient designs \cite{KimPrototype}. Rectenna nonlinearity is indeed known to be a crucial feature to be modeled and accounted for in WPT system design as it significantly influences the WPT (and SWIPT) architecture designs \cite{clerckx2018fundamentals,clerckx2016waveform}. Modeling the nonlinearity is particularly crucial in the low power regime (below 1mW), which is the primary deployment scenario of WPT in wireless networks. Importantly, the nonlinearity opens the door to more efficient signal strategies, e.g. joint waveform and beamforming, for WPT that significantly enhance the performance over beamforming-only approaches. 

\par In this paper, we derive an efficient IRS-aided WPT design by leveraging the progress made in the past few years on the modeling of the rectenna nonlinearity and its corresponding efficient signal design for WPT on one hand and on IRS on the other hand. Motivated by the significant benefits of a joint waveform and beamforming design to expand the range and the output DC power of WPT \cite{clerckx2016waveform}, \cite{huang2017large}, \cite{KimPrototype}, this work pushes the limit of IRS-aided WPT by better exploiting the presence of IRS through a joint waveform and beamforming design for both single-user and multi-user deployments. 

\par The contributions of this paper are summarized as follows.
\par \textit{First}, we design a novel architecture based on joint waveform and passive beamforming for IRS-aided WPT. We leverage the rectenna nonlinear model of \cite{clerckx2016waveform} and use it to derive a joint multisine waveform and beamforming optimization framework applicable to both single-user and multi-user IRS-aided WPT deployments. This is the first paper tackling nonlinearity and waveform design in (single-user and multi-user) IRS-aided WPT.
By leveraging jointly the frequency and spatial domains, and properly modeling the harvester nonlinearity, the proposed architecture exploits the combined benefits of the passive beamforming gain, the channel frequency selectivity and the rectenna nonlinearity. This contrasts with existing IRS-aided WPT papers \cite{pan2020intelligent,wu2019weighted,tang2019joint,wu2019joint} whose architecture only benefits from the passive beamforming gain. This also contrasts with existing papers on waveform for WPT \cite{clerckx2016waveform,huang2017large}, whose designs assume active antennas and are therefore not transferable to IRS-aided WPT due to the hardware constraints imposed by IRS. 
\par \textit{Second}, we develop two different formulations of the joint waveform and beamforming design based on a frequency-selective IRS (FS-IRS) and a frequency-flat IRS (FF-IRS). The FF-IRS assumes that the passive beamforming phases are constant across frequencies, while the FS-IRS is flexible enough to adjust the passive beamforming phases at each frequency of the multisine waveform, which  thus renders the weighted sum current problem more difficult to solve. The performance of the FS-IRS always provides an upperbound on the FF-IRS performance and helps assessing how much loss is incurred by constraining the passive beamforming phases to be the same across frequencies. This comparison sheds light on the benefits of developing more advanced IRS hardware enabling frequency-selective passive beamforming phases. This is particularly important in the context of wideband multi-band/carrier WPT since the choice of the passive beamforming phases across frequencies severely influences how much the channel frequency selectivity and the rectenna nonlinearity can be exploited by the waveform design to boost the harvested DC power. This also contrasts with existing IRS-aided WPT papers \cite{pan2020intelligent,wu2019weighted,tang2019joint,wu2019joint} that ignore such consideration since their designs assume narrowband transmissions and frequency-flat channels.  
\par \textit{Third}, under the assumption of perfect CSI, we develop an optimization framework for a general multi-user IRS-aided WPT setup based on alternating optimization (AO), successive convex approximation (SCA)  and semidefinite relaxation (SDR) to iteratively optimize the IRS passive beamforming phases and waveform weights with the others being fixed. Efficient solution is derived in two subproblems of AO by optimizing one of these variables in closed-form. Conventional WPT waveform optimization strategies mainly rely on Reverse Geometric Program (GP) \cite{clerckx2016waveform} and Semi-definite Program (SDP)\cite{huang2017waveform}, \cite{ShenCombining}, \cite{shen2020joint}.  In contrast,  for the FF-IRS, both subproblems of AO  are solved by formulating the problem into a standard semidefinite program (SDP) with SCA in an iterative manner. This is different from conventional WPT papers because the output DC current  is not only influenced by the waveform weights on each frequency (which is systematically discussed in \cite{clerckx2016waveform}) but also impacted by the passive beamforming phases in spatial and frequency domains (which is firstly promoted in this paper). In  addition, low complexity solutions are demonstrated in multi-user (MU) and single-user (SU) FS-IRS algorithms with the element-wise updating method and prior determined passive beamforming phases, respectively.
\par \textit{Fourth}, numerical results are displayed to validate the theoretical discussion on   FS-IRS and  FF-IRS WPT algorithms. Although  FS-IRS always provides an upperbound performance to  FF-IRS, FF-IRS still outperforms conventional WPT papers which emphasizes the significance of passive beamforming designs. Then, FS-IRS brings more opportunities to broadband WPT while FF-IRS is more suitable for narrowband WPT. Both  FS-IRS and FF-IRS algorithms can exhibit fast convergence which strongly reveals the effectiveness of the SCA based algorithms. Furthermore, the IRS should be deployed closed to users or BS to  obtain larger  current region. {Moreover, a near-optimal result can be achieved with low-resolution discrete-phase IRS in both FF-IRS and FS-IRS.} Numerical results confirm the inefficiency of the linear based model and demonstrate the advancement of IRS-aided WPT  on leveraging the nonlinearity of rectenna, waveform weights and passive beamforming phases.  {In addition,
this model can be extended to the multi-antenna case as in \cite{clerckx2016waveform}.}
\par $\textit{Organization}:$ In section \ref{Section_system_model}, we introduce the system model and nonlinear process of the rectenna. The joint optimization problems of  generalized MU cases  and low complexity SU scenario are discussed in section \ref{Section_multi_user}. Then, numerical results are illustrated in 
section \ref{Section_numerical_result}, followed by the conclusion and future work in section \ref{Section_conclusion}.
\par $\textit{Notations}:$ Vectors and matrices are denoted by bold and lower letters and bold and upper letters, respectively. $(\cdot)^{H}$,$(\cdot)^{\ast}$ and $(\cdot)^{T}$ denote conjugate transpose, complex conjugate and transpose. 
 ${\parallel{\cdot}\parallel}_{F}$ and ${\parallel{\cdot}\parallel}$ denote Frobenius norm and 2-norm. We use $\vert{\cdot}\vert$ for absolute value, $\text{Tr}\{\cdot\}$ for trace and $\Re\lbrace\cdot\rbrace$ and $\Im\lbrace\cdot\rbrace$ for real part operators and imaginary part operators.
The notation $\succeq$ denotes positive-semidefinite and $\mathrm{diag}\{\textbf{x}\}$ denotes diagonalization operation with elements of \textbf{x} in the diagonal. Imaginary unit is denoted as $j = \sqrt{(-1)}$. 
\section{System Model} \label{Section_system_model}
\subsection{Transmit Signal}
As is illustrated in Fig. \ref{fig: systemmodel}, we consider an IRS-aided downlink single-input-single-output (SISO) WPT system including $K$ single-antenna users, one IRS with $L$ REs and one base station (BS) equipped with single antenna at the transmitter.
The  multisine waveform at time $t$ is given as
\begin{equation}
    x(t) = \Re\biggl\{ \sum\limits_{n=1}^{N}{s}_{n}e^{j2\pi f_nt}\biggr\} = \sum\limits_{n=1}^{N}w_n\cos{(2\pi f_nt + \gamma_n)}\label{tranmit_wav}
\end{equation}
where  $s_n = w_n e^{j\gamma_n}$ ($n=1,\dots,N$) denotes the complex weight of the multisine waveform with $w_n$ and $\gamma_n$ accounting for the amplitude and phase at frequency $n$, respectively. For simplicity, we assume that the frequencies are equally spaced, i.e. $f_n = f_0 + n\Delta_f$. The transmit power constraint at the transmitter is written as $\frac{1}{2}\Vert\mathbf{s}{\Vert}^2 \leq P$ with $\mathbf{s} = [{s}_1,{s}_2,\dots,{s}_N]^T\in \mathbb{C}^{N\times 1}$ and $\frac{1}{2}$ referring to the average power of 
cosine function in (\ref{tranmit_wav}).
\vspace{-3mm}
\subsection{{Channel Model and Reflection Pattern}}
The multisine waveform propagates through wireless channels.
The channel from the BS to user $q$, from the BS to the IRS and from the  IRS to user $q$ at the frequency $n$ are
denoted as  ${h}_{\mathrm{d},q,n} $, $\mathbf{h}_{\mathrm{i},n} \in \mathbb{C}^{L\times 1}$ and $\mathbf{h}_{\mathrm{r},q,n} \in \mathbb{C}^{1\times L}$, respectively. To facilitate the reading, they are also termed as
direct channel, incident channel and reflected channel, respectively. To be specfic, 
we denote the entries of  $\mathbf{h}_{\mathrm{i},n}$ as $h_{\mathrm{i},l+(n-1)L}$  and the entries of reflected channel   $\mathbf{h}_{\mathrm{r},q,n}$ as ${h}_{\mathrm{r},q,l+(n-1)L}$ with $\mathbf{h}_{\mathrm{i},n} = [h_{\mathrm{i},1+(n-1)L},\dots,h_{\mathrm{i},L+(n-1)L}]^T$ and $\mathbf{h}_{\mathrm{r},q,n} = [{h}_{\mathrm{r},q,1+(n-1)L},\dots,{h}_{\mathrm{r},q,L+(n-1)L}]$, respectively. The channels above are assumed quasi-static and the CSIT is assumed to be perfect.

 Two IRS deployment scenarios are studied here, namely a frequency-selective IRS (FS-IRS) scenario  and a frequency-flat IRS (FF-IRS) scenario.  FS-IRS is flexible enough to adjust the passive beamforming phases in different frequencies. In contrast,  FF-IRS assumes that  passive beamforming phases are constant across frequencies.

\subsubsection{FS-IRS} The complex phasors $e^{j\psi_{l+(n-1)L}}, l = 1,2,\dots,L,$ at   frequency $n$ are collected on the diagonal entries of the diagonal matrix $\mathbf{\Theta}_n \in\mathbb{C}^{L\times L}$ with $\mathbf{\Theta}_n = \mathrm{diag}\{e^{j\psi_{1+(n-1)L}}, \dots, e^{j\psi_{L+(n-1)L}}\}, \forall n$. We assume that  $\psi_{l+(n-1)L} \in [0,2\pi)$ and $\theta_{l+(n-1)L}= e^{j\psi_{l+(n-1)L}}, l = 1,2,\dots,L$.
\subsubsection{FF-IRS} We design a common set of  passive beamforming phases that cater to all subcarriers i.e.  $\mathbf{\Theta}_n = \mathbf{\Theta} = \textrm{diag}\{e^{j\psi_ {1}}, e^{j\psi_{2}}, \dots,e^{j\psi_{L}}\},\forall n$. With $\psi_{l} \in [0,2\pi)$, we  have $\theta_{l}= e^{j\psi_{l}},\forall l$ in the remainder of this paper.
 \par
 By superposing the auxiliary channel ($\mathbf{h}_{\mathrm{r},q,n}\mathbf{\Theta}_{n}\mathbf{h}_{\mathrm{i},n}$) to the direct channel, we can obtain the  composite channel between BS and user $q$  at  frequency $n$ as
  \begin{equation}
    {h}_{q,n} = {h}_{\mathrm{d},q,n} + \mathbf{h}_{\mathrm{r},q,n}\mathbf{\Theta}_{n}\mathbf{h}_{\mathrm{i},n}. \label{composite}
  \end{equation}
{
\begin{remark}
\textit{ This paper does not deal with the implementation of FS-IRS but rather use FS-IRS as an upper bound on FF-IRS to assess how much performance could be obtained if FS-IRS could be prototyped by RF engineers. According to \cite{shen2020modeling}, the phase shift  essentially depends on the reflection coefficient of a reconfigurable impedance. Then, the frequency response of phase shift depends on the frequency response of a particular impedance. Generally, once the circuit topology  is fixed, the frequency response is fixed so that it is more complex to implement FS-IRS than FF-IRS. A practical IRS model is illustrated in \cite{li2021intelligent} which emphasizes that the amplitudes and phase of passive beamforming phases vary with the frequencies of transmitted signals. Nevertheless, there is still lack of a general model that can accurately characterize this nonlinear relationship between the practical frequency-dependent passive beamforming phases and nonlinear EH by capturing all factors. Meanwhile,  due to the different nature of systems, the objective functions and the receiver architecture, the benefits of FS-IRS over FF-IRS can be much different in WPT and in wideband OFDM communications. This means that any conclusions from IRS-aided communications may not be transferable to WPT. Hence, FF-IRS with unit modulus constraints are adopted in this paper as widely utilized in existing literature \cite{zheng2019intelligent,yang2019irs}. 
As this is the first work proposing FS-IRS and FF-IRS with nonlinear  energy harvester model 
, we would like to keep the system model as clear and simple as possible such that researchers can understand the fundamental benefits of the two proposed strategies and carry on its study in more practical model in the future.
}
\end{remark}} 
\subsection{Receive Signal}
The received signal at user $q$ of time $t$ for FS-IRS   can be expressed as\footnote{In this paper, we ignore the  signals which are reflected by IRS for two or more times due to  substantial path loss.} 
 \begin{equation}
    \begin{array}{lr}
     y_{q}(t) = 
       \Re\biggl\{\sum\limits_{n=1}^{N}({h}_{\mathrm{d},q,n}+ \mathbf{h}_{\mathrm{r},q,n}\mathbf{\Theta}_n\mathbf{h}_{\mathrm{i},n}){s}_ne^{j2\pi f_nt} \biggr\}\label{for:FS-IRSreceive}
    \end{array}
 \end{equation}
 where $q = 1,2,\dots,K$ and $y_{q}(t)$ is the input of the rectenna which is discussed in the next part.
 In FF-IRS,  $\mathbf{\Theta}_n$ is replaced with $\mathbf{\Theta},\forall n$.
\begin{figure}[t]
  \centering
    \hspace{-10mm}
    \vspace{-1mm}
    \begin{minipage}[b]{0.45\textwidth} 
    \centering 
    \includegraphics[scale = 0.45]{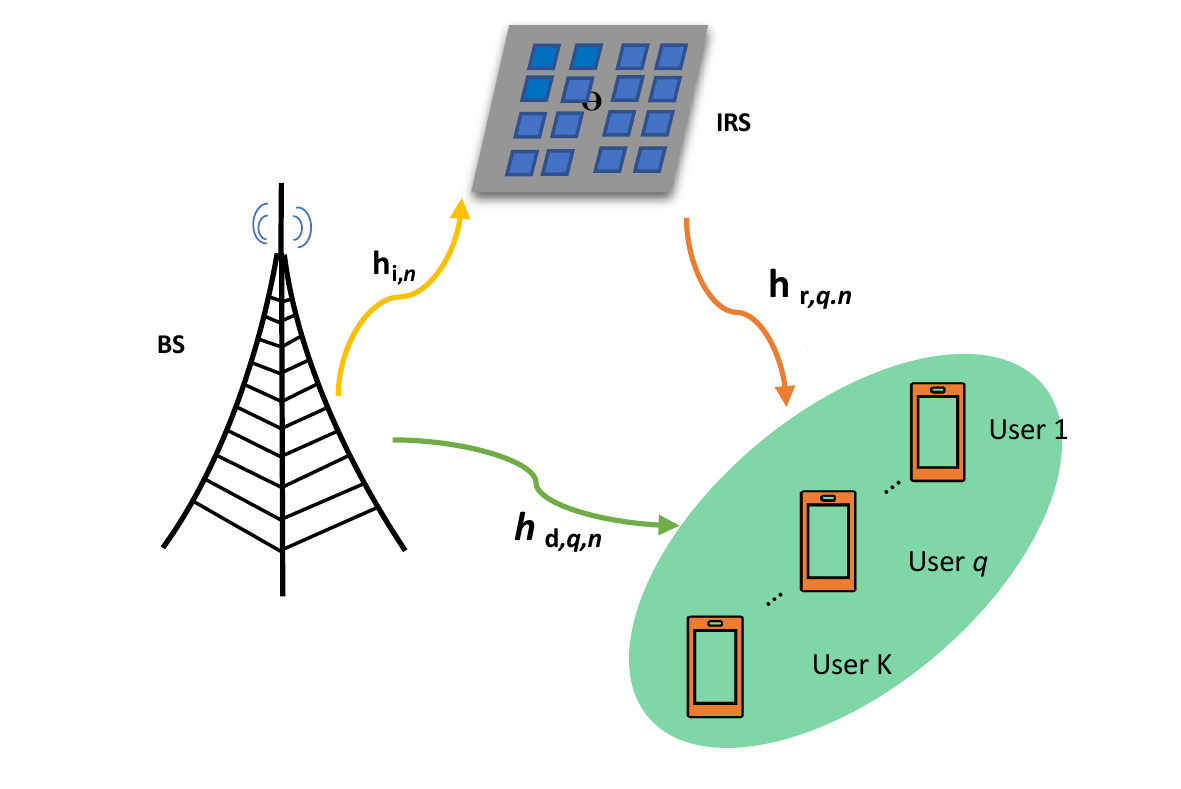}%
    \vspace{-10mm} 
  \caption{System model.}
    \label{fig: systemmodel}
    \end{minipage}
    \begin{minipage}[b]{0.45\textwidth} 
    \centering 
    \hspace{-10mm}
    \vspace{-1mm}
  \includegraphics[scale = 0.5]{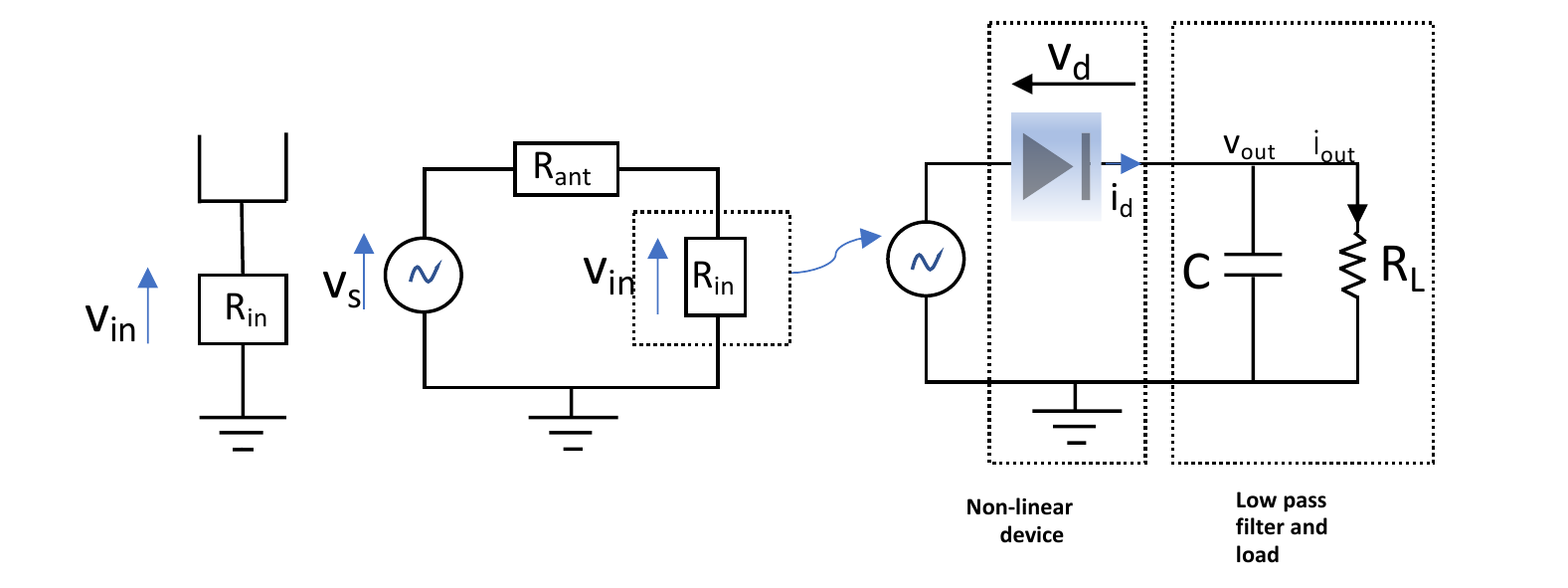}%
    \vspace{-1mm}
    \caption{Antenna equivalent circuit.}
   \label{fig: antennamodel}
   \end{minipage}
 \end{figure}
 \vspace{-3mm}
\subsection{Rectenna model}\label{subsection_rectenna}
We revisit the nonlinear process of the energy receiver in \cite{clerckx2016waveform}.
Consider a simple tractable model of the rectifier which is demonstrated in Fig. \ref{fig: antennamodel}. 
{The impinging received signal $y_q(t)$ has a average power $ P_{\mathrm{r}} = \mathcal{E}\{\vert y_{q}(t){\vert}^2\}$. The receive antenna is assumed to be a lossless antenna which can be modeled as a voltage source $v_{\textrm{s}}(t)$. Then, the rectenna architecture can be seen as a voltage source $v_\mathrm{s}{(t)}$ connected with a rectifier input impedence $Z_{\text{in}} = R_{\textrm{in}} + jX_{\textrm{in}}$ and an
antenna impedence $Z_{\textrm{ant}} = R_{\textrm{ant}} + jX_{\textrm{ant}}$.  Under the assumption of  lossless antenna and perfect matching ($R_{\textrm{ant}} = R_{\textrm{in}}, X_{\textrm{ant}} = -X_{\textrm{in}}$), the receive power $P_{\mathrm{r}}$ is completely transferred to rectenna impedence which builds the relation between $P_{\mathrm{r}}$ and $v_{\textrm{in}}(t)$ such that $   P_{\mathrm{r}} = \mathcal{E}\{\vert y_q(t){\vert}^2\} =\mathcal{E}\{\vert y_q(t){\vert}^2\}\frac{R_{\textrm{ant}}}{R_{\textrm{in}}}
  = \frac{\mathcal{E}\{\vert v_{\textrm{in}}(t){\vert}^2\}}{R_{\textrm{in}}}$.
Under the assumption of perfect matching, we have $v_{\textrm{s}}(t) = 2v_{\textrm{in}}(t)$. Hence,  $v_{\textrm{in}}(t)$ can be associated with the received signal $y_q(t)$ with  $v_{\textrm{in}}(t) = y_q(t)\sqrt{R_{\textrm{ant}}}$ \cite{clerckx2016waveform}}. 
To have a deep view of the rectifier DC output current  $i_{\textrm{out}}$,  we focus on  the diode I-V characteristic $
  i_\mathrm{d}(t) =i_\mathrm{s}(e^{\frac{v_{\textrm{in}}(t)-v_{\textrm{out}}(t)}{n^{\prime}v_\mathrm{t}}}-1) \label{for:IVdiode}
$
where $i_\mathrm{s}, n^{\prime}$ and $v_\mathrm{t}$ are the reverse bias saturation current, diode ideal factor and  thermal voltage, respectively. The diode nonlinearity is rooted in the term $e^{\frac{v_\mathrm{d}(t)}{n^{\prime}v_\mathrm{t}}}$.   \cite{clerckx2016waveform}\cite{clerckx2017wireless} illustrates the significance of the 
higher  order ($\geq 4$) truncation of the Taylor series for the term $e^{\frac{v_\mathrm{d}(t)}{n^{\prime}v_\mathrm{t}}}$ through
$i_\mathrm{d}(t) \approx \sum\limits_{i=0, \textrm{even}}^{n_0}\beta_iR_{\textrm{ant}}^{i/2} y_q(t)^i$ with $\beta_i \triangleq i_\mathrm{s}/i!(n^{\prime}v_\mathrm{t})^i$.
After the low pass filter (LPF), the  approximated output DC current $i_{\textrm{out}}$  is  given by
\begin{equation}
    i_{\textrm{out}} = \mathcal{E}\{ i_\mathrm{d}(t)\} \approx \sum\limits_{i=0, \textrm{even}}^{n_0}\beta_iR_{\textrm{ant}}^{i/2}\mathcal{E}\{ y_q(t)^i\}. \label{for:DCcurrent}
\end{equation}
To get a tractable analytical model,  one can truncate to $n_0=4$ while retaining the main source of nonlienarity as part of the fourth order term.  Higher order analysis 
can be found in \cite{clerckx2016waveform}\cite{clerckx2017wireless}. Interestingly, the 2nd order term is essentially the linear based rectenna model which is widely addressed in existed IRS-aided WPT/SWIPT papers \cite{pan2020intelligent,wu2019weighted,tang2019joint,wu2019joint} and is confirmed to be an inefficient and inaccurate in Section \ref{section_com_sufffs} {(Fig. \ref{fig: ASSN12481632M4L30}.)}.
\footnote{In addition, as a fundamental property in  I-V characteristic, the diode should operate over the nonlinear region with appropriate signal power. If the input power is too high, 
the diode will perform in linear region and (\ref{for:DCcurrent})  does not hold.}
\vspace{-4mm}
\subsection{{Discussion on Multi-Antenna Scenario}}
{The main contribution of this paper is to maximize the output DC power by leveraging the frequency diversity gain, the passive beamforming gain and the gain from EH nonlinearity. The active beamforming will not fundamentally change the observation and novelty of this paper. To keep explanation simpler without being overwhelmed by the variables of a multi-antenna scenario, multi-user SISO transmission model is chosen in our paper. However, it is easy to scale to multi-antenna system with the same AO framework in the next section and  the benefits of transmit active beamforming is shown in Fig. \ref{fig: MUK2L20Mt124P36N12481632}.}
\section{Joint design of WPT waveform and reflection elements for multi-user  scenario}\label{Section_multi_user}
 The purpose of the multi-user designs is to jointly optimize the frequency domain complex weights  of the waveform  and passive beamforming phases of IRS  to  maximize the $K$-user weighted sum output DC current,  subject to the input power constraint ($\frac{1}{2}\Vert\mathbf{s}{\Vert}^2 \leq P$) as well as the unit modulus constraints ($\left|\theta_{l+(n-1)L}\right| = 1$). 
Using the model of (\ref{for:DCcurrent}) and truncating to order 4, i.e. $n_0=4$, the approximated DC current of  user $q$ is 
$ i_{\textrm{dc},q}= \sum\limits_{n , \text{ even }, 
    n\geq 2}^{4}\beta_n R_{ant}^{n/2}\mathcal{E}\{ y_q(t)^n\}$. 
 $i_{\textrm{dc},q}$ can be found as
\begin{equation}
    i_{\textrm{dc},q} = {k}_2 \mathcal{E}\{y_q(t)^2\} + {k}_4 \mathcal{E}\{y_q(t)^4\} \label{for: initobj}
\end{equation}
where  $k_2 = \beta_2R_{\textrm{ant}} = 0.17 $ 
and $k_4 = \beta_4R_{\textrm{ant}}^2 = 957.25$ (assuming that $i_\mathrm{s} = 5\mathrm{\mu A}, {n}^{\prime}= 1.05$  and $v_\mathrm{t} = 25.86\mathrm{mV}$).
 Then, $\mathcal{E}\{y(t)^2\} $, $\mathcal{E}\{y(t)^4\}$ and the DC output current  $ i_{\textrm{dc},q}$ are calculated as
\begin{equation}
    \mathcal{E}\{y_q(t)^2\} = \frac{1}{2}\sum_{n=1}^{N}{s}_n^* {h}_{q,n}^* {h}_{q,n}{s}_n,
\end{equation}
\begin{equation}
    \begin{array}{lr}
        \mathcal{E}\{y_q(t)^4\} = 
        \frac{3}{8}\sum\limits_{\begin{subarray}{l} 
        \;\,n_1,n_2,n_3,n_4\\ n_1+n_3 = n_2 + n_4\end{subarray}}{s}_{n_1}^* {h}_{q,n_1}^* {h}_{q,n_2}{s}_{n_2}{s}_{n_3}^* {h}_{q,n_3}^* {h}_{q,n_4}{s}_{n_4},
    \end{array}
\end{equation}
\begin{equation}
    \begin{array}{lr}
    i_{\textrm{dc},q}(\mathbf{s},\{\mathbf{\Theta}_n\}_{n=1}^{N})  =  \frac{1}{2}{k}_2\sum_{n=1}^{N}{s}_n^* {h}_{q,n}^* {h}_{q,n}{s}_n +  \frac{3}{8}{k}_4 \sum\limits_{\begin{subarray}{l}
    \;\,n_1,n_2,n_3,n_4\\ n_1+n_3 = n_2 + n_4\end{subarray}}{s}_{n_1}^* {h}_{q,n_1}^* {h}_{q,n_2}{s}_{n_2}{s}_{n_3}^* {h}_{q,n_3}^* {h}_{q,n_4}{s}_{n_4},
    \end{array}\label{for: intiobj}
\end{equation}
 respectively, where the composite channel $h_{q,n}$ is given by (\ref{composite})\footnote{{When it comes to the multi-antenna scenario, the objective function (\ref{for: intiobj}) becomes $
    i_{dc}(\mathbf{s},\bm{\Theta})  =  \frac{1}{2}{k}_2\sum_{n=1}^{N}\mathbf{s}_n^H \mathbf{h}_{k,n}^H \mathbf{h}_{k,n}\mathbf{s}_n +  \frac{3}{8}{k}_4 \sum\limits_{\begin{subarray}{l} 
    n_1,n_2,n_3,n_4\\ n_1+n_3 = n_2 + n_4\end{subarray}}\mathbf{s}_{n_1}^H \mathbf{h}_{k,n_1}^H \mathbf{h}_{k,n_2}\mathbf{s}_{n_2}\mathbf{s}_{n_3}^H \mathbf{h}_{k,n_3}^H \mathbf{h}_{k,n_4}\mathbf{s}_{n_4}
$ where $\mathbf{s}_n = [s_{1+(n-1)M_t},s_{2+(n-1)M_t},\dots,s_{M_t+(n-1)M_t}]$ and $\mathbf{h}_{k,n} = \mathbf{h}_{\textrm{d}, q, n} + \mathbf{h}_{\textrm{r}, q, n}\bm{\Theta}_{n}\mathbf{H}_{\textrm{i},n} \in  \mathbb{C}^{1\times M_t}$.}}. The corresponding problem can be formulated  as a weighted sum of DC current
 \begin{maxi!}
    {\mathbf{s},\{\mathbf{\Theta}_n\}_{n=1}^{N}}  {\sum_{q= 1}^{K}\xi_q i_{\textrm{dc},q}(\mathbf{s},\{\mathbf{\Theta}_n\}_{n=1}^{N})} {\label{for:InitialMUP} }{\label{for:InitialObj}}
    \addConstraint{\Vert\mathbf{s}\Vert^2 \leq 2P}\label{for:Initialcon1}
    \addConstraint{ \vert\theta_{l+(n-1)L}\vert = 1 , n = 1,2,\dots,N,l = 1,2,\dots,L}\label{for:Initialcon2}
\end{maxi!}
where $\xi_q$ is the weight for the $q$th user.
\par
{For the alternating optimization alogrithms in the next sections, the passive beamforming $\{\mathbf{\Theta}_n^{(i)}\}_{n=1}^{N}$ and the waveform weights $\textbf{s}^{(i-1)}$ are sequentially updated in each iteration until convergence.}

\vspace{-4mm}
\subsection{Passive Beamforming  for Multi-User FF-IRS}\label{section_mu_ffirs}
In FF-IRS, we  optimize  $L$ passive beamforming phases which are constrained to be constant across all $N$ frequencies. To build this weighted sum current subproblem on  passive beamforming phases, we denote
$
    \mathbf{v}_{q,n} =\left[ \begin{matrix}   
        \textrm{diag}\{\mathbf{h}_{\mathrm{r},q,n}\}\mathbf{h}_{\mathrm{i},n}   \\
        {h}_{\mathrm{d},q,n}  
       \end{matrix}\right] \in \mathbb{C}^{(L+1)\times 1}.
$
Denote $\bm{\theta} = [\theta_{1},\theta_{2},\dots,\theta_{L},m_1] \in \mathbb{C}^{1\times (L+1) }$ with $m_1$ referring to an auxiliary variable. Then  $({h}_{\mathrm{d},q,n}+\textbf{h}_{\mathrm{r},q,n}\mathbf{\Theta}\mathbf{h}_{\mathrm{i},n}){s}_n = \bm{\theta}\mathbf{v}_{q,n}{s}_n$ and we let $\mathbf{z}_{q,n} = \mathbf{v}_{q,n}{s}_{n}$ which is grouped into $\mathbf{z}_{q} = [\mathbf{z}_{q,1}^T,\mathbf{z}_{q,2}^T,\dots,\mathbf{z}_{q,N}^T]^T \in \mathbb{C}^{(L+1)N\times 1}$. 
\begin{figure}[t]
    \centering
    \includegraphics[width = 12cm]{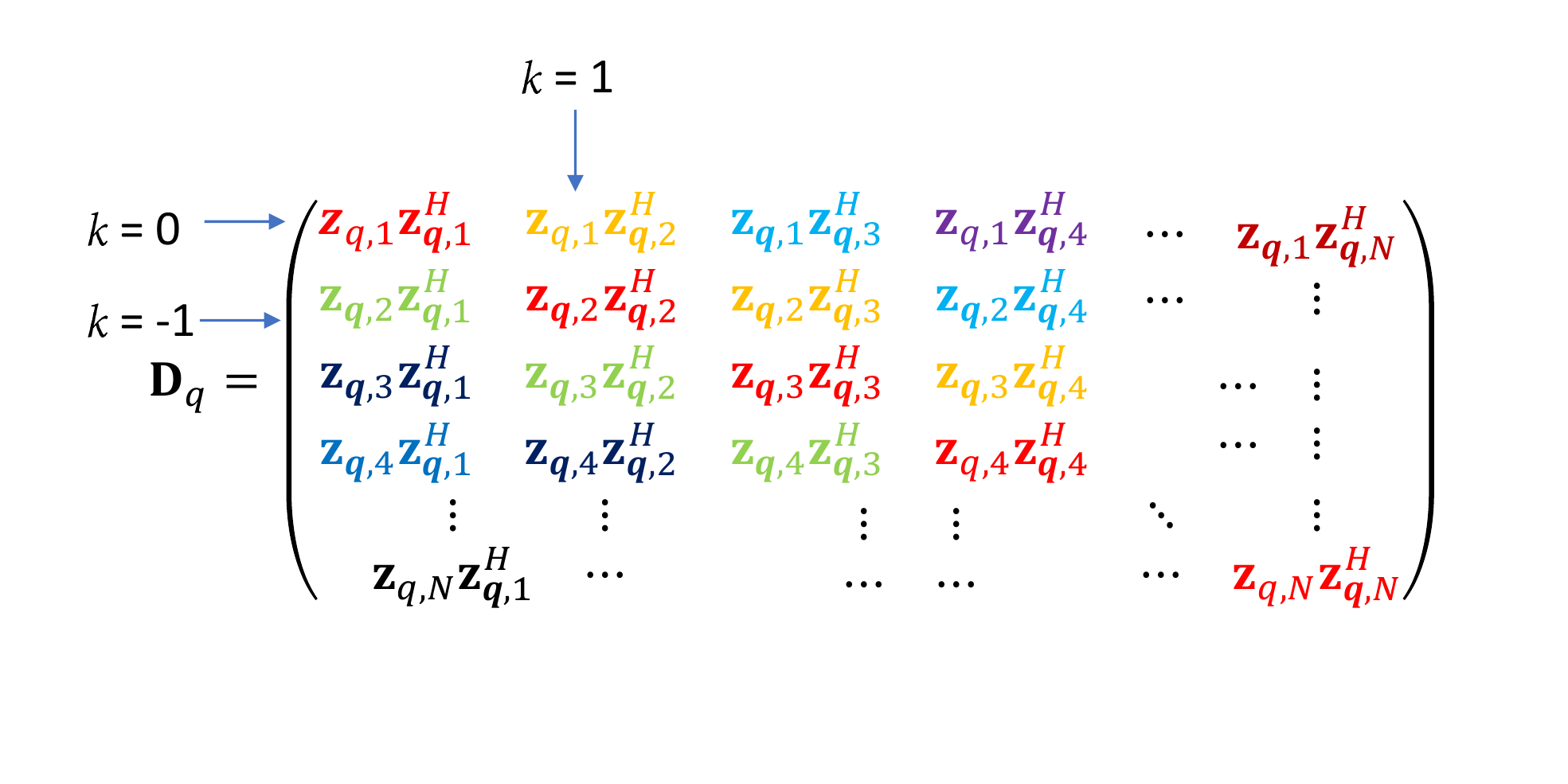}%
    \vspace{-10mm}
    \caption{$\mathbf{D}_{q,0}$ is the sum of red block diagonal, $\mathbf{D}_{q,1}$ is the sum of yellow block diagonal and $\mathbf{D}_{q,-1}$ is the sum of green block diagonal in the above matrix.}
   \label{fig:TheD}
\end{figure}
The $ i_{\textrm{dc},q}(\mathbf{s},\{\mathbf{\Theta}_n\}_{n=1}^{N})$ in (\ref{for:InitialObj})  can be equivalently rewritten as $i_{\textrm{dc},q}(\bm{\theta})$ below 
\begin{equation}
    \begin{array}{lr}
        i_{\textrm{dc},q}(\bm{\theta}) = 
        \frac{1}{2}k_2\sum\limits_{n=1}^{N}\bm{\theta}\mathbf{z}_{q,n}\mathbf{z}_{q,n}^{H}\bm{\theta}^{H}+ \frac{3}{8}k_4
        \sum\limits_{\begin{subarray}{l} 
        \,\;n_1,n_2,n_3,n_4\\ n_1+n_3 = n_2 + n_4\end{subarray}}
        \bm{\theta}\mathbf{z}_{q,n_1}\mathbf{z}_{q,n_2}^{H}\bm{\theta}^{H}\bm{\theta}\mathbf{z}_{q,n_3}\mathbf{z}_{q,n_4}^{ H}\bm{\theta}^{H}.
    \end{array}\label{for: objFFIRS}
\end{equation}
\par
To transform  (\ref{for: objFFIRS}) into a more compact form, we define 
$N(L+1)$-by-$N(L+1)$ matrix $\mathbf{D}_q$ and $(L+1)$-by-$(L+1)$ matrices $\mathbf{D}_{q,k}$
where $\mathbf{D}_q = \mathbf{z}_q\mathbf{z}_q^H$ which is shown in Fig. \ref{fig:TheD}. $k = 0$ refers to the  index of  the main block diagonal of $\mathbf{D}_q$. 
$k  \in\{1,\dots,N-1\}$ denotes the index of the $k$th  block diagonal above the main block diagonal  of the matrix $\mathbf{D}_q$.
 $k  \in\{-(N-1),\dots,-1\}$ denotes  the index of the $\vert k \vert$th block diagonal below the main block diagonal. Matrix $\mathbf{D}_{q,k}$ is obtained by summing the block matrices of the $k$th block diagonal of $\mathbf{D}_q$ e.g. 
$\mathbf{D}_{q,0} = \mathbf{z}_{q,1}\mathbf{z}_{q,1}^H + \mathbf{z}_{q,2}\mathbf{z}_{q,2}^H + \dots + \mathbf{z}_{q,N}\mathbf{z}_{q,N}^H$, 
$\mathbf{D}_{q,1} = \mathbf{z}_{q,1}\mathbf{z}_{q,2}^H + \mathbf{z}_{q,2}\mathbf{z}_{q,3}^H + \dots + \mathbf{z}_{q,N-1}\mathbf{z}_{q,N}^H$ and 
$\mathbf{D}_{q,N-1} = \mathbf{z}_{q,1}\mathbf{z}_{q,N}^H$. 
 Luckily, we can  get a compact form of (\ref{for: objFFIRS}) for the $q$th user as
\begin{equation} 
    \begin{array}{lr}
    i_{\textrm{dc},q}(\bm{\theta}) = \frac{1}{2}k_2\bm{\theta}\mathbf{D}_{q,0}\bm{\theta}^{H} + 
    \frac{3}{8}k_4\bm{\theta}\mathbf{D}_{q,0}\bm{\theta}^{H}
    (\bm{\theta}\mathbf{D}_{q,0}\bm{\theta}^{ H})^H
    + \frac{3}{4}k_4\sum\limits_{k=1}^{N-1}\bm{\theta}\mathbf{D}_{q,k}\bm{\theta}^{H}(\bm{\theta}\mathbf{D}_{q,k}\bm{\theta}^{H})^H.
    \end{array}\label{for: MUFFAOoutputidc3}
\end{equation}
Then,  we formulate weighted sum current problem which is subject to the unit modulus constraints
\begin{equation}
    \max\limits_{\bm{\theta}}  \{\sum_{q=1}^{K}\xi_q i_{\textrm{dc},q}(\bm{\theta}):\vert{\theta_l}\vert =  1, l = 1,2,\dots,L\}.
    \label{pro: ffirssimple}
\end{equation}
However, this problem   is a quartic polynomial which is NP-hard in general. To unveil this problem, we introduce auxiliary variable $d_{q,k}$ where $k\in [0,\dots,N-1]$ and
$d_{q,k}= \bm{\theta}\mathbf{D}_{q,k}\bm{\theta}^{H}$. However, for $k\neq 0$, $\mathbf{D}_{q,k}$ is non-Hermitian matrix and  the problem becomes a quadratic polynomial which is also NP-hard. To deal with this, we introduce a rank-1 positive semidefinite matrix variable $\mathbf{X} = \bm{\theta}^H\bm{\theta} \in \mathbb{C}^{(L+1)\times(L+1)}$ to linearize the problem with $d_{q,k} = \bm{\theta}\mathbf{D}_{q,k}\bm{\theta}^{H} = \mathrm{Tr}\{\mathbf{D}_{q,k}\bm{\theta}^H\bm{\theta} \} =  \mathrm{Tr}\{\mathbf{D}_{q,k}\mathbf{{X}} \}$. After defining
$\mathbf{d}_q = [d_{q,0},d_{q,1},\dots,d_{q,N-1}]^T$, $\mathbf{K}_0 = \text{diag}\{\frac{3}{8}k_4,\frac{3}{4}k_4,\dots,\frac{3}{4}k_4\}\succeq 0$ and the diagonal element of $\mathbf{X}$ as $\mathbf{X}_{l^{\prime},l^{\prime}}$ with $l^{\prime} = 1,2,\dots,L+1$,
problem (\ref{pro: ffirssimple}) can be transformed to an equivalent problem as
\begin{mini!}
    {t_0,\{\textbf{d}_q\}_{q=1}^K}{t_0}{\label{for: pff}}{\label{for: pff1}}
    \addConstraint{\sum_{q=1}^{K}\xi_q(-\frac{1}{2}k_2{d}_{q,0}-\mathbf{d}_{q}^H\mathbf{K}_0\mathbf{d}_{q})-t_0\leq 0}\label{for: pffnonconv}
    \addConstraint{\text{Tr}\{\mathbf{D}_{q,k}\mathbf{X}\} = d_{q,k} ,\forall q,k}\label{for: pff2}
     \addConstraint{\mathbf{X}_{l^{\prime},l^{\prime}} = 1, \forall l^{\prime} }\label{for: pff3}
    \addConstraint{\mathbf{X} \succeq 0}\label{for: pff4}
    \addConstraint{\textrm{rank}(\mathbf{X}) = 1}\label{for: pffrank}
\end{mini!}
which is still a  non-convex problem because of the existence of rank constraint (\ref{for: pffrank}). We apply SDR to relax (\ref{for: pffrank}) and focus on relaxed problems (\ref{for: pff1})-(\ref{for: pff4}).
As the term $\mathbf{d}_q^H\mathbf{K}_0\mathbf{d}_q$ in (\ref{for: pffnonconv}) is a non-convex quadratic constraint, SCA can be applied to address this problem to  approximate the non-convex constraint into a convex constraint and  solve the approximated convex problem iteratively. At  iteration $i$, the optimal $\mathbf{d}_q^{(i)}$  can be approximated by the optimal $\mathbf{d}_q^{(i-1)}$  via Taylor expansion \cite{adali2010adaptive}. With ${f}(\mathbf{d}_q) = \mathbf{
d}_q^H\mathbf{K}_0\mathbf{d}_q$, we have
\begin{equation}
    {f}\biggl(\mathbf{d}_q,\mathbf{d}_q^{(i-1)}\biggr) \triangleq 2\Re\biggl\{\mathbf{d}_q^{(i-1)H}\mathbf{K}_0\mathbf{d}_q\biggr\}-\mathbf{d}_q^{(i-1)H}\mathbf{K}_0\mathbf{d}_q^{(i-1)}.\label{for:SCAaffine2}
\end{equation} 
From \cite{mehanna2014feasible}, (\ref{for:SCAaffine2}) has the pattern that $ f(\mathbf{d}_q^{(i)},\mathbf{d}_q^{(i)}) =  f(\mathbf{d}_q^{(i)}) \geq f(\mathbf{d}_q^{(i)},\mathbf{d}_q^{(i-1)})$.
Accordingly, non-convex constraint (\ref{for: pffnonconv}) has the property that  $\sum_{q=1}^{K}\xi_q(-\frac{1}{2}k_2d_{q,0}-f(\mathbf{d}_q^{(i)}))\leq \sum_{q=1}^{K}\xi_q(-\frac{1}{2}k_2b_{q,0}- f(\mathbf{d}_q^{(i)},\mathbf{d}_q^{(i-1)}))\leq t_0$. Since inequation $\sum_{q=1}^{K}\xi_q(-\frac{1}{2}k_2b_{q,0}- f(\mathbf{d}_q^{(i)},\mathbf{d}_q^{(i-1)}))\leq t_0$ is a 
convex constraint at  iteration $i$, relaxed problems (\ref{for: pff1})-(\ref{for: pff4}) can be  formulated as a standard SDP as
\begin{mini!}
    {t_0,\mathbf{d}_q,\mathbf{X}\succeq 0 }{t_0}{\label{pro: sdr16}}{}
    \addConstraint{\sum_{q=1}^{K}\xi_q(-\frac{1}{2}k_2d_{q,0}- f(\mathbf{d}_q,\mathbf{d}_q^{(i-1)}))-t_0 \leq 0}\label{for: ffirsconv}
    \addConstraint{(\ref{for: pff2}), (\ref{for: pff3}) \text{ and } (\ref{for: pff4}).}
\end{mini!}
By substituting (\ref{for: pff2}) into (\ref{for: ffirsconv}), it follows that
\begin{mini!}
    {}{\text{Tr}\{\mathbf{K}_1\mathbf{X}\}}{\label{pro: 36p}}{\label{for: p16a}}
    \addConstraint{\mathbf{X}_{l^{\prime},l^{\prime}} = 1,   l^{\prime} = 1,2,\dots,L+1}\label{for: p16b}
     \addConstraint{\mathbf{X} \succeq 0}\label{for: p16c}
\end{mini!}

\noindent where   $\mathbf{K}_1 = \mathbf{J}_1 + \mathbf{J}_1^{ H}$ is a Hermitian matrix and
\begin{equation}
    \begin{array}{lr}
        \mathbf{J}_1 = \sum\limits_{q=1}^{K}\xi_q\big(-\frac{k_2}{4}\mathbf{D}_{q,0} -  \frac{3}{8}k_4d_{q,0}^{(i-1)}\mathbf{D}_{q,0} -
        \frac{3}{4}k_4\sum_{k=1}^{N-1}d_{q,k}^{(i-1)*}\mathbf{D}_{q,k}\big).
    \end{array}
\end{equation}
This standard SDP problem (\ref{pro: 36p}) can be solved by invoking existing softwares, e.g., CVX MATLAB \cite{grant2014cvx}.
Denoting the solution as $\textbf{X}^{\star}$, if rank($\textbf{X}^{\star}) = 1$, the SDR is tight and $\textbf{X}^{\star}$ is a stationary point of (\ref{pro: 36p}) so that a local optimal solution can be extracted by $\mathbf{X}^{\star} =\bm{\theta}^{\star(i)}\bm{\theta}^{H\star(i)}$ \cite{luo2010semidefinite}. If  rank($\textbf{X}^{\star}) > 1$, we can only use the Gaussian randomization method in \cite{ma2004semidefinite,ma2002quasi} to extract a  suboptimal rank-1 solution $\bm{\theta}^{\prime(i)} = [\theta_1^{\prime(i)},\theta_2^{\prime(i)},\dots,\theta_L^{\prime(i)},m_1^{(i)}]$, retrieve the phase shifts by $\bm{\theta}^{\star(i)} = \bm{\theta}^{\prime(i)}/m_1^{(i)}$ and group the first $L$ elements into the diagonal of $\mathbf{\Theta}^{\star(i)} = \textrm{diag}\{\bm{\theta}^{\star(i)}[1:L]\}$. Then, we have the  composite  channel ${h}_{q,n} = ({h}_{\mathrm{d},q,n} + \mathbf{h}_{\mathrm{r},q,n}\mathbf{\Theta}^{\star(i)}\mathbf{h}_{\mathrm{i},n})$. 
{
\begin{remark}
\textit{Note that different from \cite{pan2020intelligent} where a predefined minimum harvested power threshold is addressed, problem (\ref{pro: 36p}) is a standard SDP in each iteration with only unit modulus constraint. 
The known approximation accuracy for this complex constant modulus problem is $\pi/4$ which guarantees that a good approximation can be found if we have a rank-1 solution  \cite{luo2010semidefinite}. When we evaluated the results in Section \ref{Section_numerical_result}, we found that all channel realizations  generated a rank-1 $\mathbf{X}^{\star}$. This is particularly important since Algorithm \ref{alg:MU-FF-IRS} is guaranteed to provide a stationary point for problem (\ref{for:InitialMUP}) with a rank-1 solution in the simulations (see Appendix \ref{appendixB}). Even though the performance is degraded by the  Gaussian randomization method later, the performance loss is still  negligible if   $\mathbf{X}^{\star}$ is rank-1 for all tested channels \cite{wu2018intelligent}.}
\end{remark}}
\vspace{-3mm}
\subsection{Passive Beamforming  for Multi-User FS-IRS}\label{section_mu_fsirs}
In MU FS-IRS, different from problem  (\ref{pro: ffirssimple}), we aim to maximize the weighted sum current  subject to  $NL$  unit modulus constraints and the problem can be written as
\begin{equation}
    \max\limits_{\{\mathbf{\Theta}_n\}_{n=1}^{N}}  \{\sum_{q=1}^{K}\xi_q i_{\textrm{dc},q}(\{\mathbf{\Theta}_n\}_{n=1}^{N}):\vert{\theta_{l+(n-1)L}}\vert =  1,  n \in 1,2,\dots,N , l \in 1,2,\dots,L\}.
    \label{pro: fsirssimple}
\end{equation}
To obtain a compact form , we define
$
    \mathbf{v}_{q,n} =\left[ \begin{matrix}   
        \textrm{diag}\{\mathbf{h}_{\mathrm{r},q,n}\}\mathbf{h}_{\mathrm{i},n}   \\
        {h}_{\mathrm{d},q,n}  
       \end{matrix}\right] \in \mathbb{C}^{(L+1)\times 1}.
$ We
denote $\bm{\theta}_n = [\theta_{1+(n-1)L},\dots,\theta_{L+(n-1)L},m_n] \in \mathbb{C}^{1\times (L+1) },\forall n$ with $m_n$ referring to an auxiliary variable.  We have $({h}_{\mathrm{d},q,n}+\mathbf{h}_{\mathrm{r},q,n}\mathbf{\Theta}_n\mathbf{h}_{\mathrm{i},n}){s}_n = \bm{\theta}_n\mathbf{v}_{q,n}{s}_n$.  Denote $\mathbf{z}_{q,n} = \mathbf{v}_{q,n}{s}_{n}\in \mathbb{C}^{(L+1)\times 1}$ which is grouped into $\mathbf{z}_{q} = [\mathbf{z}_{q,1}^T,\mathbf{z}_{q,2}^T,\dots,\mathbf{z}_{q,N}^T]^T \in \mathbb{C}^{(L+1)N\times 1}$.
(\ref{for: intiobj}) can be rewritten as
\begin{equation}
    \begin{array}{lr}
         i_{\textrm{dc},q}(\{\bm{\theta}_n\}_{n=1}^{N}) = \frac{1}{2}k_2\sum_{n=1}^{N}\bm{\theta}_n\mathbf{z}_{q,n}\mathbf{z}_{q,n}^{H}\bm{\theta}^{H}_n+ 
         \frac{3}{8}k_4\sum\limits_{\begin{subarray}{l}\,\;n_1,n_2,n_3,n_4\\ n_1+n_3 = n_2 + n_4\end{subarray}}
    \bm{\theta}_n\mathbf{z}_{q,n_1}\mathbf{z}_{q,n_2}^{H}\bm{\theta}^{H}_n\bm{\theta}_n\mathbf{z}_{q,n_3}\mathbf{z}_{q,n_4}^{ H}\bm{\theta}^{H}_n.
    \end{array}\label{obj_muirsin}
\end{equation}
\par
To formulate (\ref{obj_muirsin}) into a more tractable compact form, we denote $\bm{\theta} = [\bm{\theta}_1,\dots,\bm{\theta}_N]\in \mathbb{C}^{1\times N(L+1) }$ and define  $(L+1)N$-by-$(L+1)N$ matrices $\mathbf{E}_{q}$ and $\mathbf{E}_{q,k}$ with $\mathbf{E}_{q} = \mathbf{z}_{q}\mathbf{z}_{q}^H$.  $k = 0$ denotes the index of the main block diagonal in $\mathbf{E}_{q}$. $k\in \{1,\dots,N-1\}$ denotes the index of the $k$th block diagonal above main diagonal of $\mathbf{E}_{q}$.  $k\in\{-(N-1),\dots,-1\}$ denotes the index of the $\vert k\vert$th block diagonal below the main block diagonal.  $\mathbf{E}_{q,k}$ is obtained by retaining the $k$th block diagonal of $\mathbf{E}_{q}$ and setting other block matrices as $\mathbf{0}_{(L+1)\times (L+1)}$.  Then, the  output current for user $q$ can be equivalently written as
\begin{equation} 
    \begin{array}{lr}
    i_{\textrm{dc},q}(\bm{\theta}) = \frac{1}{2}k_2\bm{\theta}\mathbf{E}_{q,0}\bm{\theta}^{H} + 
    \frac{3}{8}k_4\bm{\theta}\mathbf{E}_{q,0}\bm{\theta}^{H}(\bm{\theta}\textbf{E}_{q,0}\bm{\theta}^{H})^H
    + \frac{3}{4}k_4\sum\limits_{k=1}^{N-1}\bm{\theta}\mathbf{E}_{q,k}\bm{\theta}^{H}(\bm{\theta}\mathbf{E}_{q,k}\bm{\theta}^{H})^H.
    \end{array}\label{for: MUFSIRSp}
     \end{equation}
To linearize the quartic function (\ref{for: MUFSIRSp}), we take $\bm{\theta}\mathbf{E}_{q,k}\bm{\theta}^{H}= e_{q,k}$  which is collected into 
$\mathbf{e}_q = [e_{q,0},e_{q,1},\dots,e_{q,N-1}]^T$. (\ref{for: MUFSIRSp}) can be written as
$
        i_{\textrm{dc},q}(\mathbf{e}_{q})
        = \frac{1}{2}k_2e_{q,0} + \mathbf{e}_{q}^H\mathbf{K}_0\mathbf{e}_{q}.\label{for: MUIRSp1}
$
The problem can be formulated as
\begin{mini!}
    {t_1,{\{\mathbf{e}_{q}\}}_{q=1}^{K}}{t_1}{\label{for: p81}}{}
    \addConstraint{\sum\limits_{q=1}^{K}\xi_q(-\frac{1}{2}k_2e_{q,0}-\mathbf{e}_{q}^H\mathbf{K}_0\mathbf{e}_{q})-t_1\leq 0}\label{for: p82}
    \addConstraint{\bm{\theta}\mathbf{E}_{q,k}\bm{\theta}^{H}= e_{q,k} ,\forall q,k}\label{for: p83}
    \addConstraint{\left|{\theta}_{l+(n-1)L}\right| = 1, \forall n,l.}\label{for: p85}
\end{mini!}
\par
We still approximate the non-convex constraint (\ref{for: p82}) iteratively by the SCA which is similar with that in Section \ref{section_mu_ffirs} and the problem can be reformulated as
\begin{equation}
    \begin{array}{lr}
         \min\limits_{{\bm{\theta}}}\quad {\bm{\theta}\mathbf{K}_2\bm{\theta}^{H}}  \\
        \text{ s.t. } \quad(\ref{for: p83})  \text{ and }(\ref{for: p85})\label{for: p91}
    \end{array}
\end{equation}
where  $\mathbf{K}_2 = \mathbf{J}_2+\mathbf{J}_2^{H}$ is a Hermitian matrix and $\mathbf{J}_2$ is
\begin{equation}
    \begin{array}{lr}
        \mathbf{J}_2= \sum\limits_{q=1}^{K}\xi_q\big(-\frac{k_2}{4}\mathbf{E}_{q,0} - \frac{3}{8}k_4e_{q,0}^{(i-1)}\mathbf{E}_{q,0}
    - \frac{3}{4}k_4\sum_{k=1}^{N-1}[e_{q,k}^{(i-1)}]^*\mathbf{E}_{q,k}\big).
    \end{array}
\end{equation}
If we apply the same approach as in Section \ref{section_mu_ffirs} to problem (\ref{for: p91}), that is, by denoting  $\mathbf{X} = \bm{\theta}^H\bm{\theta}$ and formulating problem (\ref{for: p91}) into a SDP, solving this intricate SDP will render a complexity around  $\textrm{O}((NL)^6)$ which
demonstrates low feasibility for large number of frequencies. 
To tackle this issue,  here, we illustrate a low complexity strategy to optimize the $NL$ elements in $\bm{\theta}$ with a complexity around $\textrm{O}((NL)^2)$.
\par
\textit{Element-Wise Updating Method}:
 Element-Wise Updating Method (EWU)  sequentially optimizes  $N(L+1)$ variables ($NL$ passive beamforming phases variables and $L$ auxiliary variables) in $\bm{\theta}$  which is also applied in \cite{wu2019beamforming} and \cite{guo2019weighted}. To facilitate the reading, in this method, we denote $\mathbf{K} = -\mathbf{K}_2$   and  the element at the $i$th row and the $j$th column of $\mathbf{K}$ as $k_{i,j}, i = 1,2,\dots, N(L+1), j = 1,2,\dots, N(L+1)$.
In addition, we denote the $m$th variable in $\bm{\theta}$  as $\theta_m$ in this part for simplicity (instead of $\theta_{l + (n-1)L}$ in the preceding discussion).  minimization of 
${\bm{\theta}\mathbf{K}_2\bm{\theta}^{H}}$ in problem (\ref{for: p91}) becomes maximization of $\bm{\theta}\mathbf{K}\bm{\theta}^H$ with the same unit modulus constraints (\ref{for: p85}). Since $\theta_m$ is  updated  with other $N(L+1) - 1$ elements in $\bm{\theta}$ being fixed, $\bm{\theta}\mathbf{K}\bm{\theta}^H$ can be written as an element-wise function $f(\theta_m)$ below
\begin{equation}
    \begin{array}{lr}
        f({\theta_m}) =  \theta_mk_{m,m}\theta_m^* + \sum\limits_{j=1,j\neq m}^{N(L+1)}\theta_mk_{m,j}\theta_j^* +
         \sum\limits_{i=1,i\neq m}^{N(L+1)}\theta_ik_{i,m}\theta_m^*+\sum\limits_{i\neq m,j\neq m}\theta_ik_{i,j}\theta_j\label{for: EUW3}
    \end{array}
\end{equation}
where $m = 1,\dots, N(L+1)$.
\par
Due to the fact that $\mathbf{K}$ is a Hermitian matrix, we substitute $k_{i,j} = k_{j,i}^*$ and yield
\begin{equation}
    \begin{array}{lr}
        f({\theta_m}) = \theta_mk_{m,m}\theta_m^* + 2\Re\{\sum\limits_{j=1,j\neq m}^{N(L+1)}\theta_mk_{m,j}\theta_j^*\} +
        \sum\limits_{i\neq m,j\neq m}\theta_ik_{i,j}\theta_j.
    \end{array}
\end{equation}
In order to optimize $\theta_m$, we abandon other irrelevant elements. The objective function $g(\theta_m)$ becomes
\begin{equation}
    g(\theta_m) = k_{m,m}\vert\theta_m^*\vert^2 + 2\Re\{\sum\limits_{j=1,j\neq m}^{N(L+1)}\theta_mk_{m,j}\theta_j^*\}.\label{for: EUW1}
\end{equation}
Since $\vert\theta_m^*\vert^2 = 1$, maximizing $g(\theta_m)$ is to maximize the term $\Re\{\sum\limits_{j=1,j\neq m}^{N(L+1)}\theta_mk_{m,j}\theta_j^*\}$. 
After denoting $\sum\limits_{j=1,j\neq m}^{N(L+1)}k_{m,j}\theta_j^*  = C_{m}$, the optimal $\theta_m^{\star}$ is
\begin{equation}
    \angle\theta_m^{\star} = -\angle C_{m}, \theta_m^{\star} = e^{-j\angle C_{m}}.\label{for: EUW2}
\end{equation}
Therefore, $\theta_m^{\star}$ can be determined one by one until all $N(L+1)$ variables are updated.
\par After this SCA based EWU method, similarly, the composite channel for user $q$ is obtained by grouping the first $L$ elements of $\bm{\theta}_n$ into the diagonal matrix $\mathbf{\Theta}_n^{\star(i)} = \mathrm{diag}\{\bm{\theta}_n^{\star(i)}[1:L]\}$ and substituting the $\mathbf{\Theta}_n^{\star(i)}$ into the auxiliary channel to get ${h}_{q,n} = {h}_{\mathrm{d},q,n}+\mathbf{h}_{r,q,n}\mathbf{\Theta}_n^{\star(i)}\mathbf{h}_{\mathrm{i},n}$.
\subsection{Waveform Design for  Multi-User FF-IRS and Multi-User FS-IRS}\label{section_wdmuirs}
\begin{figure}[t]
    \centering
    \includegraphics[width=12cm]{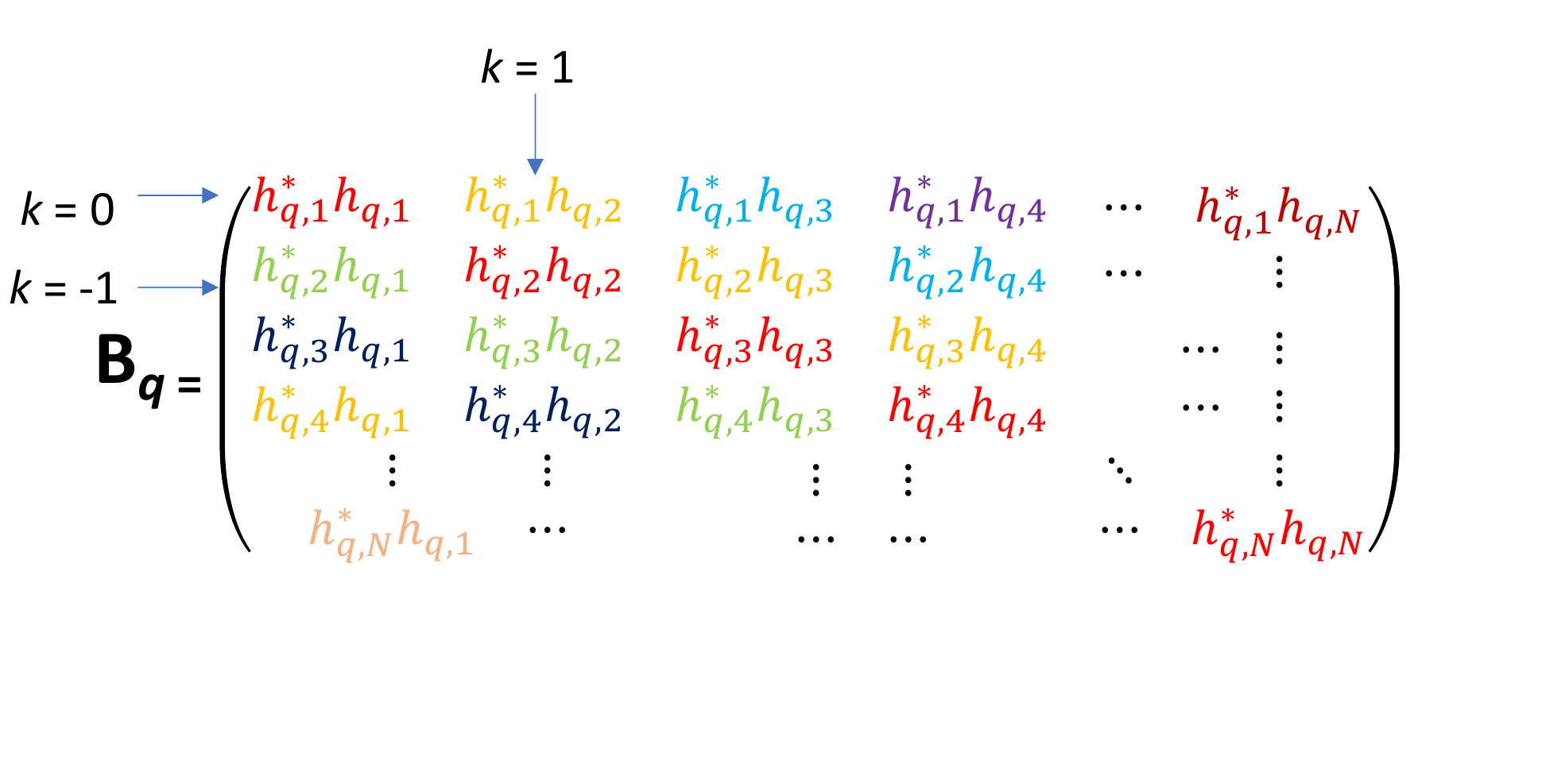}
    \vspace{-15mm}
    \caption{The definition of $\mathbf{B}_q$. }\label{Fig:TheBK}
\end{figure}

In this section, we aim to maximize the weighted sum current  subject to the power constraint at the transmitter for the composite channel $h_{q,n}$  obtained in the previous sections by optimizing the waveform weights at different frequencies. 
To formulate this subproblem into a compact form, we introduce   $N$-by-$N$ matrices 
$\mathbf{B}_{q} = \mathbf{h}_q^H\mathbf{h}_q$ with $\mathbf{h}_q = [h_{q,1},h_{q,2},\dots,h_{q,n}]$, which is shown in  Fig. \ref{Fig:TheBK}. 
$k  \in\{1,\dots,N-1\}$ denotes  the index of the $k$th  diagonal above the main diagonal (with index $k = 0$) of the matrix $\mathbf{B}_q$.  $k  \in\{-(N-1),\dots,-1\}$ denotes the index of the $\vert k \vert$th  diagonal below the main  diagonal.  $\mathbf{B}_{q,k}$ is obtained by retaining the $k$th  diagonal of $\mathbf{B}_q$ with other entries being zeros. The compact form of the output current at the $q$th user $i_{\textrm{dc},q}(\mathbf{s})$ can be recast as
\begin{equation} 
    \begin{array}{lr}
    i_{\mathrm{dc},q}(\mathbf{s}) = \frac{1}{2}k_2\mathbf{s}^H\mathbf{B}_{q,0}\mathbf{s} + \frac{3}{8}k_4\mathbf{s}^H\mathbf{B}_{q,0}\mathbf{s}(\mathbf{s}^H\mathbf{B}_{q,0}\mathbf{s})^H
    + \frac{3}{4}k_4\sum\limits_{k=1}^{N-1}\mathbf{s}^H\mathbf{B}_{q,k}\mathbf{s}(\mathbf{s}^H\mathbf{B}_{q,k}\mathbf{s})^H. \label{for:MUFFAOoutputidc2}
    \end{array}
\end{equation}
\par
The  weighted sum current of all users can be written as $\sum\limits_{q=1}^{K}\xi_qi_{\textrm{dc},q}(\mathbf{s})$.
Denote the auxiliary variable $b_{q,k} = \mathbf{s}^H\mathbf{B}_{q,k}\mathbf{s}  = \mathrm{Tr}\{\mathbf{B}_{q,k}\mathbf{s}\mathbf{s}^H \} = \mathrm{Tr}\{\mathbf{B}_{q,k}\mathbf{Y} \}$ for $q = 1,\dots,K, k = 0,\dots,N-1$, such that $\mathbf{b}_q = [b_{q,0},\dots,b_{q,N-1}]^T$.  We have $i_{\textrm{dc}}(\{\mathbf{b}_q\}_{q=1}^{K}) =
 \sum_{q=1}^{K}\xi_q(\frac{1}{2}k_2b_{q,0}+\mathbf{b}_q^H\mathbf{K}_0\mathbf{b}_q)$. The problem becomes
\begin{mini!}
    {t_2,\{\mathbf{b}_q\}_{q=1}^K}{t_2}{\label{for:p61}}{\label{for:p6}}
    \addConstraint{\sum_{q=1}^{K}\xi_q(-\frac{1}{2}k_2b_{q,0}-\mathbf{b}_{q}^H\mathbf{K}_0\mathbf{b}_{q})-t_2\leq 0}\label{for:p62}
    \addConstraint{\mathrm{Tr}\{\mathbf{B}_{q,k}\mathbf{Y}\} = b_{q,k} ,\forall q,k}\label{for:p63}
    \addConstraint{ \mathrm{Tr}\{\mathbf{Y}\} \leq 2P}\label{for:p65}
    \addConstraint{\textrm{rank}\{\mathbf{Y}\} = 1.}\label{for:p66}
\end{mini!}
Similar to the  approach in Section \ref{section_mu_ffirs}, we  relax the rank constraint (\ref{for:p66}) and use the SCA to solve it.
Let ${g}(\mathbf{b}_q,\mathbf{b}_q^{(i-1)}) = 2\Re\{\mathbf{b}_q^{(i-1)H}\mathbf{K}_0\mathbf{b}_q\} - \mathbf{b}_q^{(i-1)H}\mathbf{K}_0\mathbf{b}_q^{(i-1)}$. We have $-\frac{1}{2}k_2b_{q,0}-{g}(\mathbf{b}_q^{(i)},\mathbf{b}_q^{(i)}) \leq -\frac{1}{2}k_2b_{q,0}-{g}(\mathbf{b}_q^{(i)},\mathbf{b}_q^{(i-1)})\leq t_2.$
Therefore, $-\frac{1}{2}k_2b_{q,0}-{g}(\mathbf{b}_q,\mathbf{b}_q^{(i-1)})  \leq t_2$ can be seen as an upper bound convex constraint of initial non-convex constraint (\ref{for:p62}).  The problem can be rewritten as
\begin{mini!}
    {{t}_2,\mathbf{b}_q,\mathbf{Y}\succeq 0}{t_2 }{}{}
    \addConstraint{-\frac{1}{2}k_2b_{q,0}-{g}(\mathbf{b}_q,\mathbf{b}_q^{(i-1)}) -t_2 \leq 0}
    \addConstraint{(\ref{for:p63}) \textrm{ and } (\ref{for:p65})}
\end{mini!} 
which can be written into a SDP as
\begin{equation}
    \begin{array}{lr}
       \min\limits_{\mathbf{Y}}\{\mathrm{Tr}\{\mathbf{K}_3\mathbf{Y}\}:\mathrm{Tr}\{\mathbf{Y}\} \leq 2P\} \label{for:p8}
    \end{array}
\end{equation}
where   $\mathbf{K}_3= \mathbf{J}_3+ \mathbf{J}_3^{H}$ is a Hermitian matrix and 
\begin{equation}
    \begin{array}{lr}
        \mathbf{J}_3 = \sum\limits_{q=1}^{K}\xi_q\big(-\frac{k_2}{4}\mathbf{B}_{q,0} - \frac{3}{8}k_4b_{q,0}^{(i-1)}\mathbf{B}_{q,0}
    - \frac{3}{4}k_4\sum_{k=1}^{N-1}b_{q,k}^{(i-1)*}\mathbf{B}_{q,k}\big).
    \end{array}
\end{equation}
\subsection{Alternating Optimization}
For all the channel trails, we iteratively update the passive beamforming phases and waveform weights until convergence, which is shown in Algorithm \ref{alg:MU-FF-IRS} and Algorithm \ref{alg:MU-FS-IRS}. In both algorithms,  $[\mathbf{V}_{\mathbf{K}_3}]_{\min}$ refers to the eigenvector  corresponding to the smallest eigenvalue which
can be obtained   via the eigenvalue decomposition (EVD) \cite{mehanna2014feasible}. 
\begin{algorithm}[t]
    \caption{MU FF-IRS WPT Algorithm}\label{alg:MU-FF-IRS}
    \begin{algorithmic}[1]
        \State{$\textbf{Initialize}$:  $i= 0$,  $\{\mathbf{d}_q^{(0)}$,$\mathbf{b}^{(0)}_q$,$\mathbf{B}_{q}^{(0)}$,$\mathbf{D}_q^{(0)}\}_{q=1}^{K}$,$\mathbf{K}_1^{(0)}$,$\mathbf{K}_3^{(0)}$,$\mathbf{\Theta}^{(0)}$, $\mathbf{y}^{(0)}$ and corresponding $t_2^{(0)}$;}
        \State{$\textbf{Repeat}$}
        \State{$i = i+1$;}               
        \State{Update $\mathbf{K}_1^{(i)}$ and compute $\bm{\theta}^{\star(i)}$ using  Gaussian randomization method in \cite{ma2004semidefinite}, $\mathbf{\Theta}^{(i)\star}= \textrm{diag}\{\bm{\theta}^{\star(i)}(1:L)\};$}
        \State{Update $\mathbf{B}_{q,k}^{(i)} \textrm{ and  } b_{q,k}^{(i)}, \forall q,k$;}
        \State{Update $\mathbf{K}_3^{(i)}$, get eigenvector $\mathbf{y}^{\star} = \sqrt{2P}[\mathbf{V}_{\mathbf{K}_3}]_{\min};\mathbf{Y}^{\star} = \mathbf{y}^{\star} \mathbf{y}^{\star H}$;}
        \State{Update $\mathbf{Y}^{(i)} = \mathbf{Y}^{\star}$. Then update $d_{q,k}^{(i)},\mathbf{D}_{q,k}^{(i)} ,\forall q,k$, compute $t_2^{(i)}$;}
        \State{$\textbf{Until } \vert t_2^{(i)} - t_2^{(i-1)}\vert/{\vert t_2^{(i)}\vert}\leq \varepsilon$ or $i\geq i_{\max}$}
        \State{$\mathbf{s}^{\star} = \mathbf{y}^{\star}$, output $\mathbf{\Theta}^{\star}$.}
    \end{algorithmic}
\end{algorithm}
\begin{algorithm}[t]
    \caption{MU FS-IRS WPT Algorithm}\label{alg:MU-FS-IRS}
    \begin{algorithmic}[1]
        \State{$\textbf{Initialize}$:  $i= 0$, $\{\mathbf{e}_q^{(0)}$,$\mathbf{b}^{(0)}_q$,$\mathbf{B}_{q}^{(0)}$,$\mathbf{E}_q^{(0)}\}_{q=1}^{K}$,$\mathbf{K}_2^{(0)}$, $\mathbf{K}_3^{(0)}$, $\{\mathbf{\Theta}^{(0)}_n\}_{n=1}^{N}$, $\mathbf{y}^{(0)}$ and corresponding $t_2^{(0)}$;}
        \State{$\textbf{Repeat}$}
        \State{$i = i+1$;}               
        \State{Update $\mathbf{K}_2^{(i)}$ and compute $\bm{\theta}^{*(i)}$ using EWU  and generate $\{\mathbf{\Theta}_n^{(i)\star}\}_{n=1}^{N}= \{\textrm{diag}\{\bm{\theta}_n^{\star(i)}(1:L)\}\}_{n=1}^{N}$;}
        \State{Update $\mathbf{B}_{q,k}^{(i)} \text{ and  } b_{q,k}^{(i)}, \forall q,k$;}
        \State{Update $\mathbf{K}_3^{(i)}$, get eigenvector $\mathbf{y}^{\star} = \sqrt{2P}[\mathbf{V}_{\mathbf{K}_3}]_{\min};\mathbf{Y}^{\star} = \mathbf{y}^{\star} \mathbf{y}^{\star H}$;}
        \State{Update $\mathbf{Y}^{(i)} = \mathbf{Y}^{\star}$ and $e_{q,k}^{(i)},\mathbf{E}_{q,k}^{(i)} ,\forall q,k$ and compute $t_2^{(i)}$;}
        \State{$\textbf{Until } \vert t_2^{(i)} - t_2^{(i-1)}\vert/{\vert t_2^{(i)}\vert}\leq \varepsilon$ or $i\geq i_{\max}$}
        \State{$\mathbf{s}^{\star} = \mathbf{y}^{\star}$, output  $\{\mathbf{\Theta}_n^{\star}\}_{n=1}^{N}$.}
    \end{algorithmic}
\end{algorithm}
\subsection{Convergence}
\begin{prop}\label{Proposition_alg2_phase}
{For any feasible initial points,  Algorithm  \ref{alg:MU-FF-IRS} can converge and provide a $\textbf{X}^{\star}$ which satisfies the Karush-Kuhn-Tucker (KKT) conditions of problem (\ref{pro: 36p}), although there is no guarantee $\textbf{X}^{\star}$ is rank-1.}
\end{prop}
\par $Proof$: 
{See Appendix \ref{appendixA}.}
$\hfill\Box$
\begin{prop}\label{Proposition_alg1_phase}
{If the $\textbf{X}^{\star}$ in \textit{Proposition} \ref{Proposition_alg2_phase} is rank-1, $\textbf{X}^{\star}$ is also the local optimal solution of problem (\ref{for: pff}). Then, Algorithm \ref{alg:MU-FF-IRS} converges to a local optimal solution in (\ref{for:InitialMUP}).}
\end{prop}
\par $Proof$: {See Appendix \ref{appendixB}.}
$\hfill\Box$
\begin{prop}\label{Proposition_alg3_phase}
{For any feasible initial points,  the passive beamforming vectors $\bm{\theta}^{\star}$ generated by Algorithm  \ref{alg:MU-FS-IRS} in problem (\ref{for: p91}) can guarantee the non-decreasing property of $\sum_{q=1}^{K}\xi_q i_{\textrm{dc},q}(\bm{\theta}^{\star})$. The  Algorithm \ref{alg:MU-FS-IRS} can converge to a suboptimal point of (\ref{for:InitialMUP}).}
\end{prop}
\par$Proof$: The SCA firstly ensures the non-decreasing property as $\bm{\theta}^{\star(i-1)}$ is also a feasible solution in iteration $i$. In the EWU method, (\ref{for: EUW1}) and (\ref{for: EUW2}) double guarantee the non-decreasing procedure by applying the complementary phasor of $\sum\limits_{j=1,j\neq m}k_{m,j}\theta_j^*$ in (\ref{for: EUW1}). {Due to the unit modulus constraint, the SCA based EWU strategy is upper bounded and finally converge. The Algorithm \ref{alg:MU-FS-IRS} is suboptimal due to the performance loss incurred by the EWU method. This performance loss is demonstrated in Fig. \ref{fig: alg2alg3compareK1L102030P36N12481632}.}
$\hfill\Box$
\begin{prop}\label{Proposition_alg23_waveform}
For any feasible initial points, the waveform design subproblem in Section \ref{section_wdmuirs} can converge to the stationary points of problem (\ref{for:p61}) with $\sum_{q=1}^{K}\xi_q( i_{\textrm{dc},q}(\mathbf{s}^{\star(i-1)}) \leq \sum_{q=1}^{K}\xi_q( i_{\textrm{dc},q}(\mathbf{s}^{\star(i)})$.
\end{prop}
\par$Proof$: See \cite{huang2017large} for detail.
$\hfill\Box$
\subsection{Low Complexity Design for Single-user FS-IRS Scenario}\label{Section_single_user}
In this section,  we demonstrate an extra  algorithm for SU FS-IRS for three reasons. {First, Algorithm \ref{alg:MU-FS-IRS} can only provide  suboptimal solutions due to  the EWU strategy, but Algorithm \ref{alg:SU-FS-IRS}  exhibits global optimal solutions. Second, Algorithm \ref{alg:SU-FS-IRS}  further reduces the complexity compared with Algorithm \ref{alg:MU-FS-IRS} with $K = 1$ without  the necessity of AO  strategy.  Third, Algorithm \ref{alg:MU-FS-IRS} can not boil down to Algorithm \ref{alg:SU-FS-IRS}  when $K = 1$ since, in Algorithm \ref{alg:MU-FS-IRS}, perfect alignment between the auxiliary channel and the direct channel may not be achieved,  which incurs a performance loss compared with  Algorithm \ref{alg:SU-FS-IRS}}. This performance loss is demonstrated in Fig. \ref{fig: alg2alg3compareK1L102030P36N12481632}.
\par
Because of the issues above, we demonstrate an efficient algorithm for SU FS-IRS by priorly determining the  passive beamforming phases  and waveform phases. For simpler notation,  ${h}_{\mathrm{d},q,n}$ and $\mathbf{h}_{\mathrm{r},q,n}$    boil down to ${h}_{\mathrm{d},n}$ and $ \mathbf{h}_{\mathrm{r},n}$   respectively. We denote ${h}_{\mathrm{d},n} = A_{\mathrm{d},n}e^{\phi_{\mathrm{d},n}}$, ${h}_{\mathrm{i},l+(n-1)L} = A_{\mathrm{i},l+(n-1)L}e^{\phi_{\mathrm{i},l+(n-1)L}}$ and  ${h}_{\mathrm{r},l+(n-1)L} = A_{\mathrm{r},l+(n-1)L}e^{\phi_{\mathrm{r},l+(n-1)L}}$. Then, we denote the entries of  $\mathbf{A}_{\mathrm{i},n}$ as $A_{\mathrm{i},l+(n-1)L}$ and the entries of  $\mathbf{A}_{\mathrm{r},n}$ as  ${A}_{\mathrm{r},l+(n-1)L}$ with $\mathbf{A}_{\mathrm{i},n} = [A_{\mathrm{i},1+(n-1)L},\dots,A_{\mathrm{i},L+(n-1)L}]^T$ and $\mathbf{A}_{\mathrm{r},n} = [{A}_{\mathrm{r},1+(n-1)L},\dots,{A}_{\mathrm{r},L+(n-1)L}]$, respectively.
The passive beamforming and frequency domain power allocation for SU FS-IRS are sequentially demonstrated in the next sections.
\subsubsection{Passive Beamforming for Single-User FS-IRS}
\par
Since  FS-IRS is assumed to have an independent reflection for all frequencies with totally $NL$ variables. When $K$
 = 1, each passive beamforming phase in FS-IRS can idealy align the auxiliary channel to the direct channel and
the optimal passive beamforming phases and waveform phases have the values below
 \begin{equation}
    \gamma_{n}^{\star} = -\phi_{\mathrm{d},n}, \forall n, \label{for:phaseofsignal}
 \end{equation}
 \begin{equation}
    \psi_{l+(n-1)L}^{\star} = -(\gamma_{n}^{\star} +  \phi_{\mathrm{i},l+(n-1)L} +  \phi_{\mathrm{r},l+(n-1)L}), \forall n,l. \label{for:phaseofRE}
 \end{equation}
\begin{remark}
\textit{Recall that in Section \ref{section_wdmuirs}, we directly optimize the waveform weight $\mathrm{s}_n$ which consists of the amplitude $w_n$  and phase $\gamma_n$. In contrast, due to the prior determination of $\gamma_n$, the joint passive beamforming and waveform design in SU FS-IRS  is essentially   the frequency domain power allocation at different frequencies, i.e. $w_n$,  and naturally exhibits lower complexity  than SCA based AO strategy in   MU FS-IRS. }
\end{remark}
\subsubsection{Frequency Domain Power Allocation for Single-User FS-IRS}
\par Due to the ideal alignment of composite channel in SU FS-IRS, we denote the corresponding composite channel $A_n =  {A}_{\mathrm{d},n}+\mathbf{A}_{r,n}\mathbf{A}_{\mathrm{i},n}$ which is collected into  $\mathbf{A} =  [{A}_{1},\dots, {A}_{N}]^T$ and group $w_n$ into the power allocation vector $\mathbf{p} = [w_1,w_2,\dots,w_N]^T$. Objective function (\ref{for: intiobj}) is transformed to
\begin{equation} 
    \begin{array}{lr}
    i_{\textrm{dc}}(\mathbf{p}) = \frac{1}{2}k_2\sum_{n=1}^{N} w_n^2{A}_n^2 + 
     \frac{3}{8}k_4\sum\limits_{\begin{subarray}{m} 
    \,\;n_1,n_2,n_3,n_4\\ n_1+n_3 = n_2 + n_4\end{subarray}}w_{n_1}{A}_{n_1}
{A}_{n_2} w_{n_2}w_{n_3}{A}_{n_3}{A}_{n_4} w_{n_4}.\label{for:idcoutputMRT}
    \end{array}
\end{equation}
\par
We can  formulate (\ref{for:idcoutputMRT}) into a more compact form by introducing $N$-by-$N$ matrices $\mathbf{B}$ and $\mathbf{B}_{k}$ where $\mathbf{B} = \mathbf{A}\mathbf{A}^T$.
$k  \in\{1,\dots,N-1\}$ denotes the index of  the $k$th  diagonal above the main diagonal (with the index $k = 0$) of the matrix $\mathbf{B}$. $k  \in\{-(N-1),\dots,-1\}$ denotes the index of the $\vert k \vert$th  diagonal below the main  diagonal.  $\mathbf{B}_{k}$ is obtained by retaining the $k$th  diagonal of $\mathbf{B}$ with other entries as zeros. The compact form of (\ref{for:idcoutputMRT}) can be expressed as
 \begin{equation}
    \begin{array}{lr}            
        i_{\textrm{dc}}(\mathbf{p}) = \frac{1}{2}k_2\mathbf{p}^H\mathbf{B}_0\mathbf{p} + \frac{3}{8}k_4\mathbf{p}^H\mathbf{B}_0\mathbf{p}(\mathbf{p}^H\mathbf{B}_0\mathbf{p})^H
        + \frac{3}{4}k_4\sum\limits_{k=1}^{N-1}\mathbf{p}^H\mathbf{B}_k\mathbf{p}(\mathbf{p}^H\mathbf{B}_k\mathbf{p})^H.\label{for:FSoutputidccomp}
   \end{array}
 \end{equation}
 To maximize the output DC current under the power constraint at transmitter, the problem can be formulated as
\begin{equation}
    \max\limits_{\mathbf{p}}  \{ i_{\textrm{dc}}(\mathbf{p}):\Vert\mathbf{p}\Vert^2\leq 2P\}.\label{pro: first}
\end{equation}
Similar to that in Section \ref{section_wdmuirs}, we introduce a rank-1 positive semidefinite matrix variable $\mathbf{P} = \mathbf{p}\mathbf{p}^H$ to linearize the problem with $b_k = \mathbf{p}^H\mathbf{B}_k\mathbf{p} = \mathrm{Tr}\{\mathbf{B}_k\mathbf{{P}} \}$. Then,
denote $\mathbf{b} = [b_0,b_1,\dots,b_{N-1}]^T$. Finally, we have
$i_{\textrm{dc}}(\mathbf{b}) = \frac{1}{2}k_2b_0+\frac{3}{8}k_4b_0b_0^*+\frac{3}{4}k_4\sum_{k=1}^{N-1}b_kb_k^* = \frac{1}{2}k_2b_0+\mathbf{b}^H\mathbf{K}_0\mathbf{b}$. The problem (\ref{pro: first}) can be equivalently written  as
\begin{mini!}
    {t_3,\mathbf{b},\mathbf{P}\succeq 0}{t_3}{\label{for:p21}}{}
    \addConstraint{-i_{\textrm{dc}}(\mathbf{b}) - t_3 \leq 0 }\label{for:p22}
    \addConstraint{\mathrm{Tr} \{\mathbf{B}_k\mathbf{P}\} = b_k, \forall k}\label{for:p23}
    \addConstraint{\mathrm{Tr} \{\mathbf{P}\} \leq 2P}\label{for:p25}
    \addConstraint{\textrm{rank}\{\mathbf{P}\} = 1}\label{for:p26}.
\end{mini!}
After  the relaxation of rank constraint and  SCA approach,
the compact form of problem (\ref{pro: first}) can be formulated as
\begin{equation}
    \min\limits_{\mathbf{P}\succeq 0 }\{\text{Tr}\{\mathbf{K}_4\mathbf{P}\}:\text{Tr}\{ \mathbf{P}\}\leq 2P\}\label{for:relaxk4}
\end{equation}
where  $\mathbf{K}_4 = \mathbf{J}_4 + \mathbf{J}_4^{ H} $ is a Hermitian matrix and $\mathbf{J}_4 = -\frac{k_2}{4}\mathbf{B}_0- \frac{3}{8}k_4b_0^{(i-1)}\mathbf{B}_0
   - \frac{3}{4}k_4\sum_{k=1}^{N-1}b_{k}^{(i-1)*}\mathbf{B}_k.$ The whole procedure is summarized in Algorithm \ref{alg:SU-FS-IRS}.
\begin{remark}
\textit{According to \cite{huang2017large}, problem (\ref{for:relaxk4}) has a  rank-1 global optimal solution  $\mathbf{P}^{\star} = \mathbf{p}^{\star}\mathbf{p}^{\star H}$ with $\mathbf{p}^{\star} = \sqrt{2P}[\mathbf{V}_{\mathbf{K}_4}]_{\min}$, which is different from that in generalized MU scenario as {the EWU} based strategy can only provide   suboptimal solutions. In each iteration, the rank-1  global optimal solution   not only guarantees to demonstrate convergence to the stationary points of the relaxed problem (\ref{for:relaxk4}) but also the original problem (\ref{for:p21}). }
\end{remark}
\begin{algorithm}[t]
    \caption{SU FS-IRS WPT Algorithm}\label{alg:SU-FS-IRS}
    \begin{algorithmic}[1]
        \State{$\textbf{Initialize}$: $i= 0$ and $\mathbf{b}^{(0)}$, $\mathbf{p}^{(0)}$ and $t_3^{(0)};$}
        \State{$\textbf{Repeat}$}
            \State{$i = i+1$}   
            \State{Compute $\mathbf{K}_4  = \mathbf{J}_4 + \mathbf{J}_4^{ H}$, get eigenvector $\mathbf{p}^{\star} = \sqrt{2P}[\mathbf{V}_{\mathbf{K}_4}]_{\min};\mathbf{P}^{\star} = \mathbf{p}^{\star} \mathbf{p}^{\star H}$;}
            \State{Update $\mathbf{P}^{(i)} = \mathbf{P}^{\star}$; Update $b_k^{(i)} = \mathrm{Tr}\{\mathbf{B}_k\mathbf{P}^{(i)} \}, \forall k$;}
            \State{Update $t_3^{(i)}$;}
            \State{$\textbf{Until } \vert t_3^{(i)} - t_3^{(i-1)}\vert/{\vert t_3^{(i)}\vert}\leq \varepsilon$ or $i\geq i_{\max}$}
        \State{${s}_n^{\star} = [\mathbf{p}^{\star}]_{n,1}\cdot {h}_n^{*}/\vert{h}_n\vert, \forall n$ .}
    \end{algorithmic}
\end{algorithm}
\subsection{{Discussion on Large Scale Scenario of SU FS-IRS}}
\begin{figure}[t]
    \centering
    \begin{minipage}[b]{0.45\textwidth} 
    \centering 
    \includegraphics[width = 7cm]{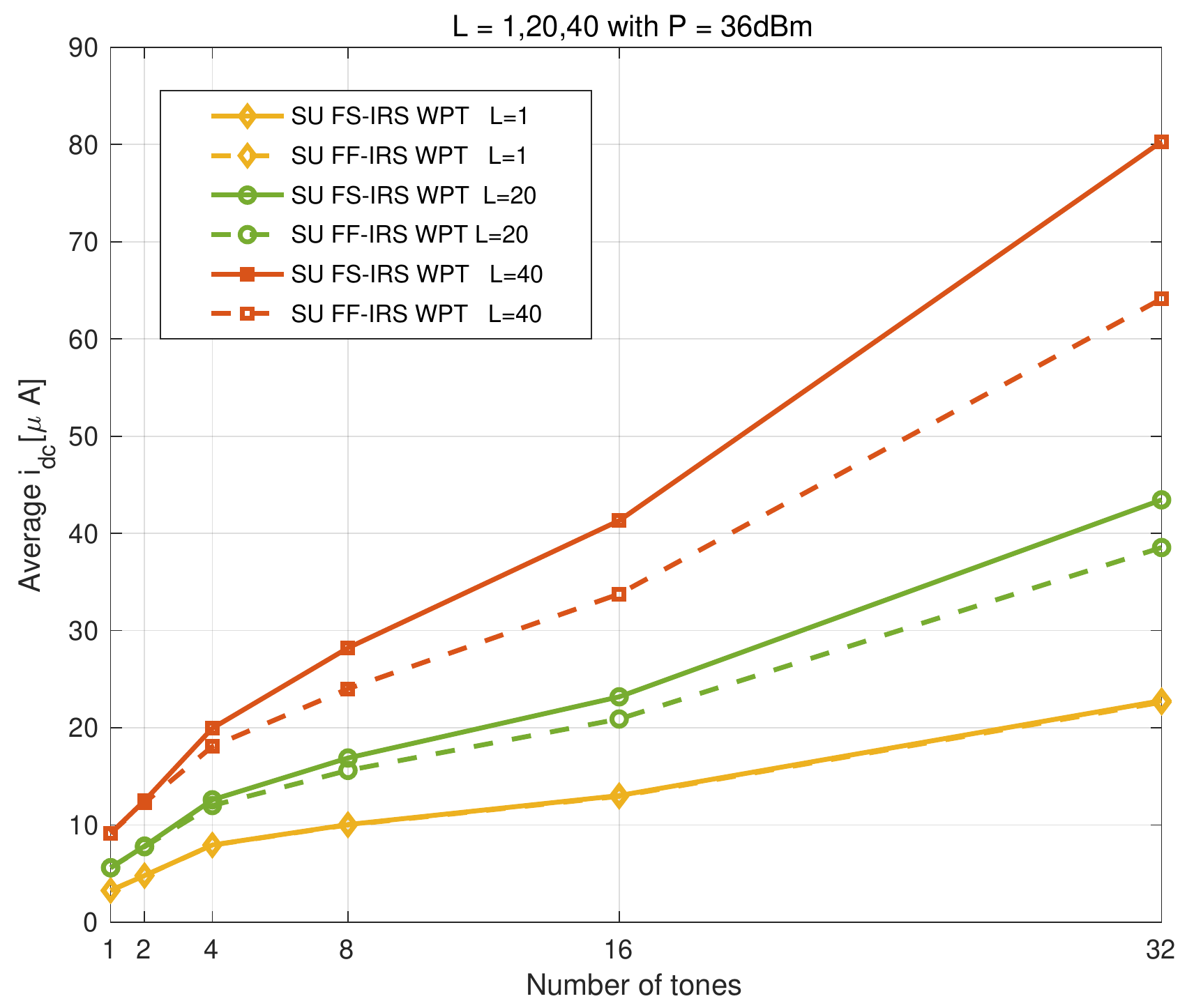}
     \vspace{-6mm}
    \caption{Average $i_{dc}$ versus $N$ for $L = 1,20,40$.}
   \label{fig: L11020Mt5}
    \end{minipage}
     \vspace{10mm}
    \begin{minipage}[b]{0.45\textwidth} 
    \centering 
    \includegraphics[width = 7.1cm,height = 5.95cm]{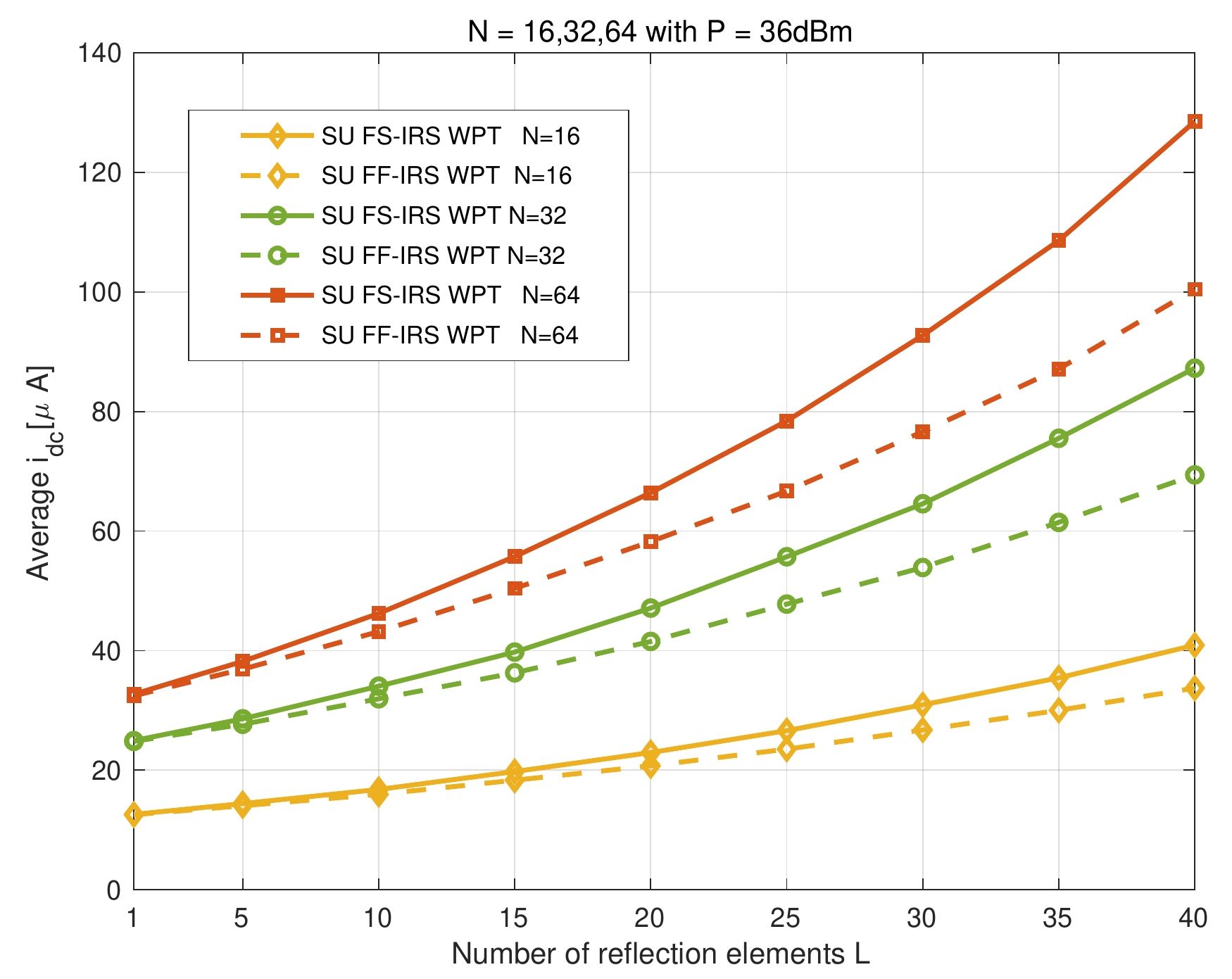}
     \vspace{-6mm}
    \caption{Average $i_{dc}$ versus $L$ for $N = 16,32,64$.}
   \label{fig: L151015202530Mt1N64}
    \end{minipage}
     \vspace{10mm}
   \begin{minipage}[b]{0.45\textwidth} 
    \centering 
    \includegraphics[width = 7.2cm]{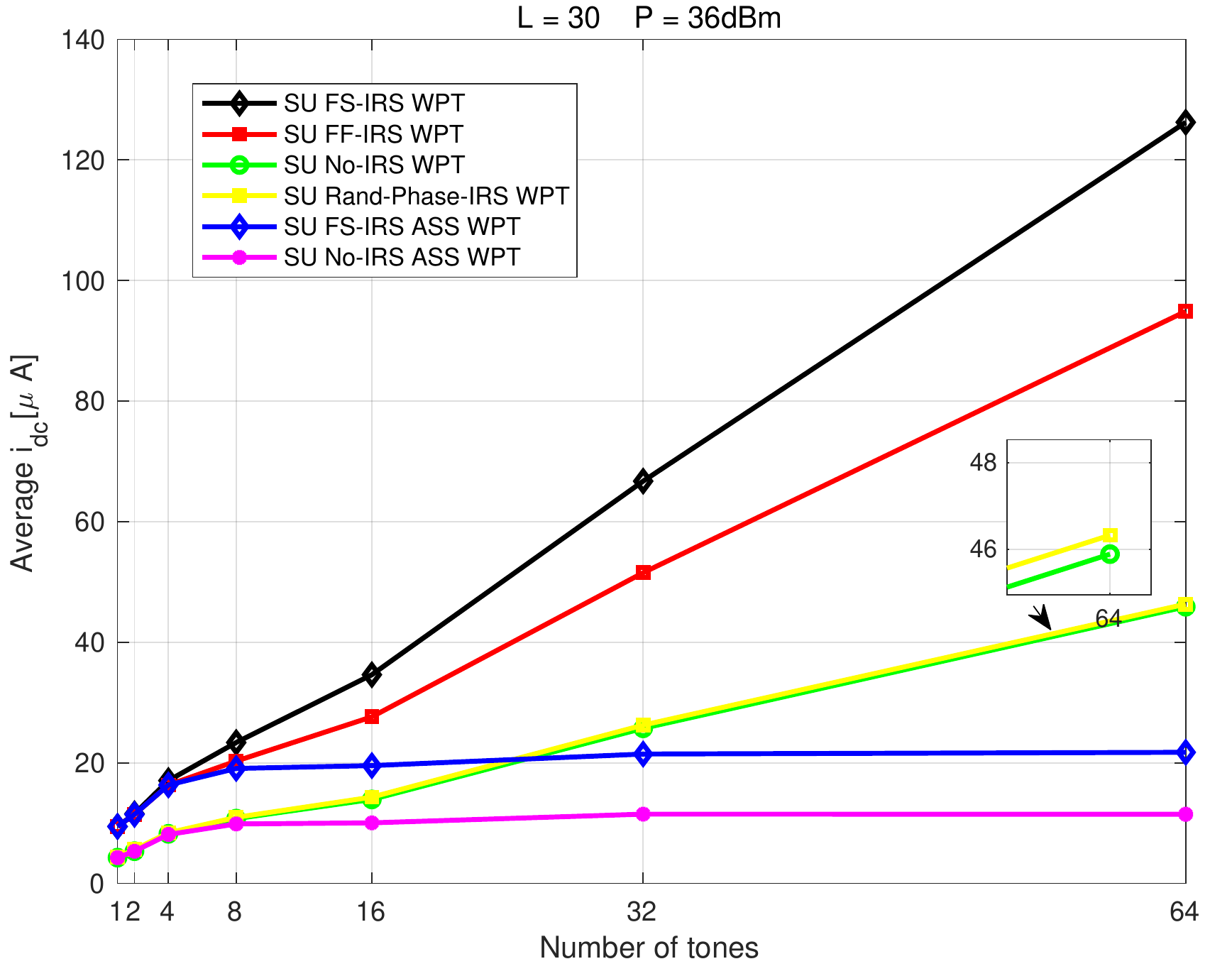}
    \vspace{-8mm}
    \caption{Average $i_{dc}$ versus $N$.}
   \label{fig: ASSN12481632M4L30}
    \end{minipage}  
   \begin{minipage}[b]{0.45\textwidth} 
    \centering
   
    \includegraphics[width = 7.7cm, height = 6.2cm]{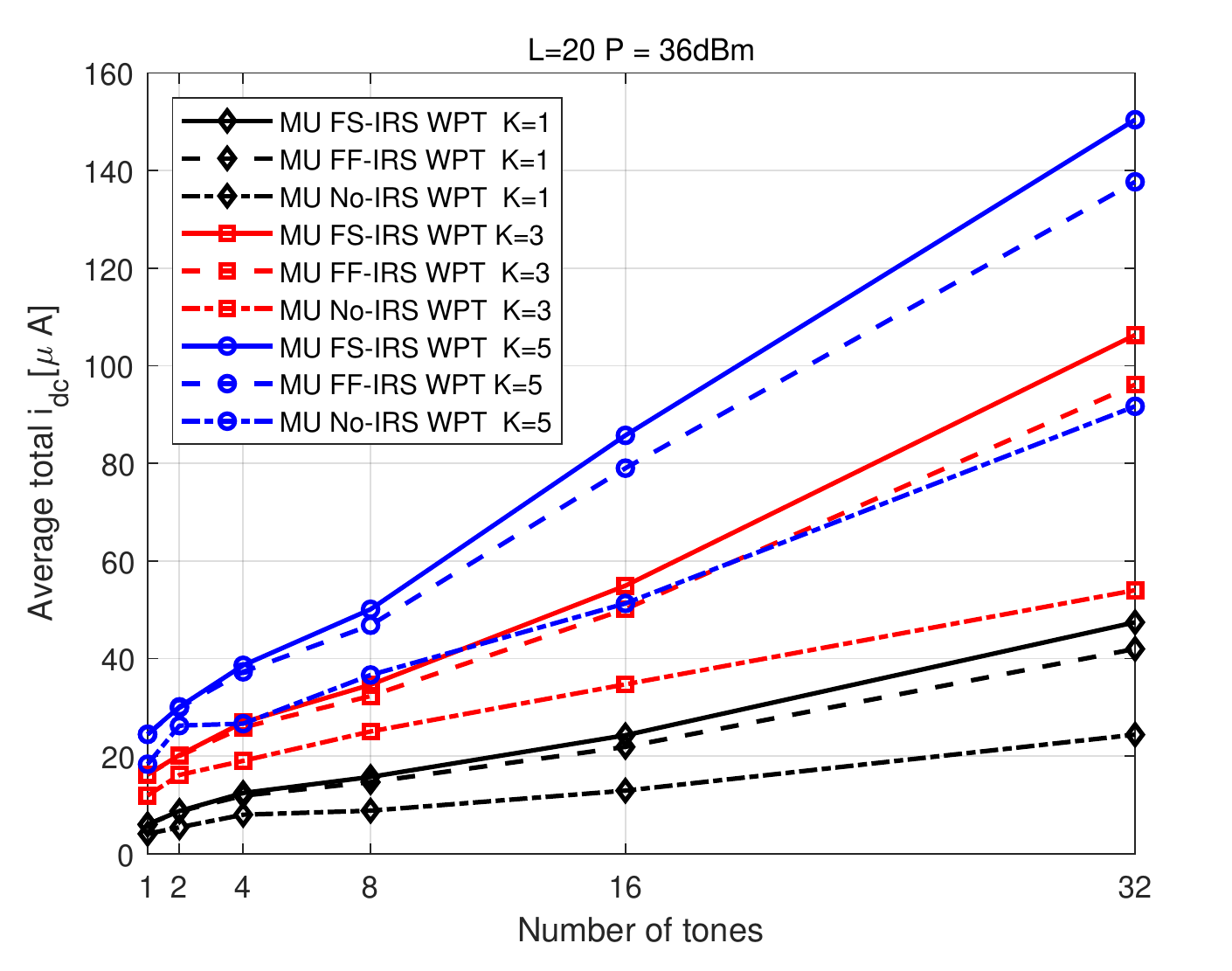}
     \vspace{-18mm}
    \caption{The average weighted sum of $i_{\textrm{dc}}$ versus $N$ for $K = 1,3,5$ and $L = 20$.}
   \label{fig: MUK135compare}
   \end{minipage}
\end{figure}

{This subsection proposes a brief analysis of the performance of SU FS-IRS in the limit of a large number of REs $L$ and frequencies $N$. We assume that all the channels are uncorrelated. Given the large scale fading for ${h}_{\mathrm{d},n}, h_{\mathrm{i}, l + (n-1)L}$ and $h_{\mathrm{r},  l + (n-1)L}$ as $\Lambda_{\mathrm{d}}^{1/2}, \Lambda_{\mathrm{i}}^{1/2}$ and $\Lambda_{\mathrm{r}}^{1/2}$, respectively,  under the perfect alignment in SU FS-IRS, we have
 \vspace{-1mm}
\begin{equation}
    \begin{array}{lr}
       \mathcal{E}\{\vert{h}_n\vert\} = \mathcal{E}\bigg\{\sqrt{\vert h_{\mathrm{d},n} + 
        \sum\limits_{l=1}^{L}{h}_{\mathrm{r},l+ (n-1)L}\theta_{l+(n-1)L}{h}_{\mathrm{i},l+(n-1)L}{\vert}^2}\bigg\}
         \overset{L\rightarrow \infty}{\approx} \sqrt{\Lambda_{\mathrm{d}} + \Lambda_\mathrm{i}\Lambda_\mathrm{r}L^2}.\label{for:effectH}
    \end{array}
\end{equation}
According to \cite{huang2017large}, with uniform power
allocation across frequencies, we can approximate the output current as
\vspace{-2mm}
\begin{equation}
    \begin{array}{lr}
        i_{dc} \overset{N\rightarrow \infty}{\approx} k_2P(\Lambda_{\mathrm{d}}+\Lambda_\mathrm{i}\Lambda_\mathrm{r}L^2) + 
        \frac{3}{2}k_4P^2(\Lambda_{\mathrm{d}}+\Lambda_\mathrm{i}\Lambda_\mathrm{r}L^2)^2 + 3k_4 P^2(\Lambda_{\mathrm{d}}+\Lambda_\mathrm{i}\Lambda_\mathrm{r}L^2)^2N.\label{large}
    \end{array}
    \vspace{-1mm}
\end{equation}}
{Therefore, if the number of $L$ is large, the output current will 
scale up with the square number of $L^2$ (i.e. $L^4$) which is shown in (\ref{large}) and validated in Fig. \ref{fig: L151015202530Mt1N64}. This is due to the nonlinearity of the rectenna model which is truncated to $n_0 = 4$. In addition, if $N$ is large enough, the output DC current linearly scales with $N$ which is validated  in Section \ref{Section_numerical_result}\footnote{{Based on the rectenna model in Section \ref{subsection_rectenna}, increasing the number of subcarriers improves the harvested DC power as long as the EH does not operates in the saturation region. In a practical prototype of No-IRS WPT, due to the Peak-to-Average Power Ratio (PAPR)  limits of the transmitter, 16 is a good number which was experimentally observed in \cite{KimPrototype,kim2020range}.}}. This is inline with the scaling law in \cite{clerckx2016waveform}. }

\begin{figure}[t]
    \centering
    \includegraphics[scale = 0.6]{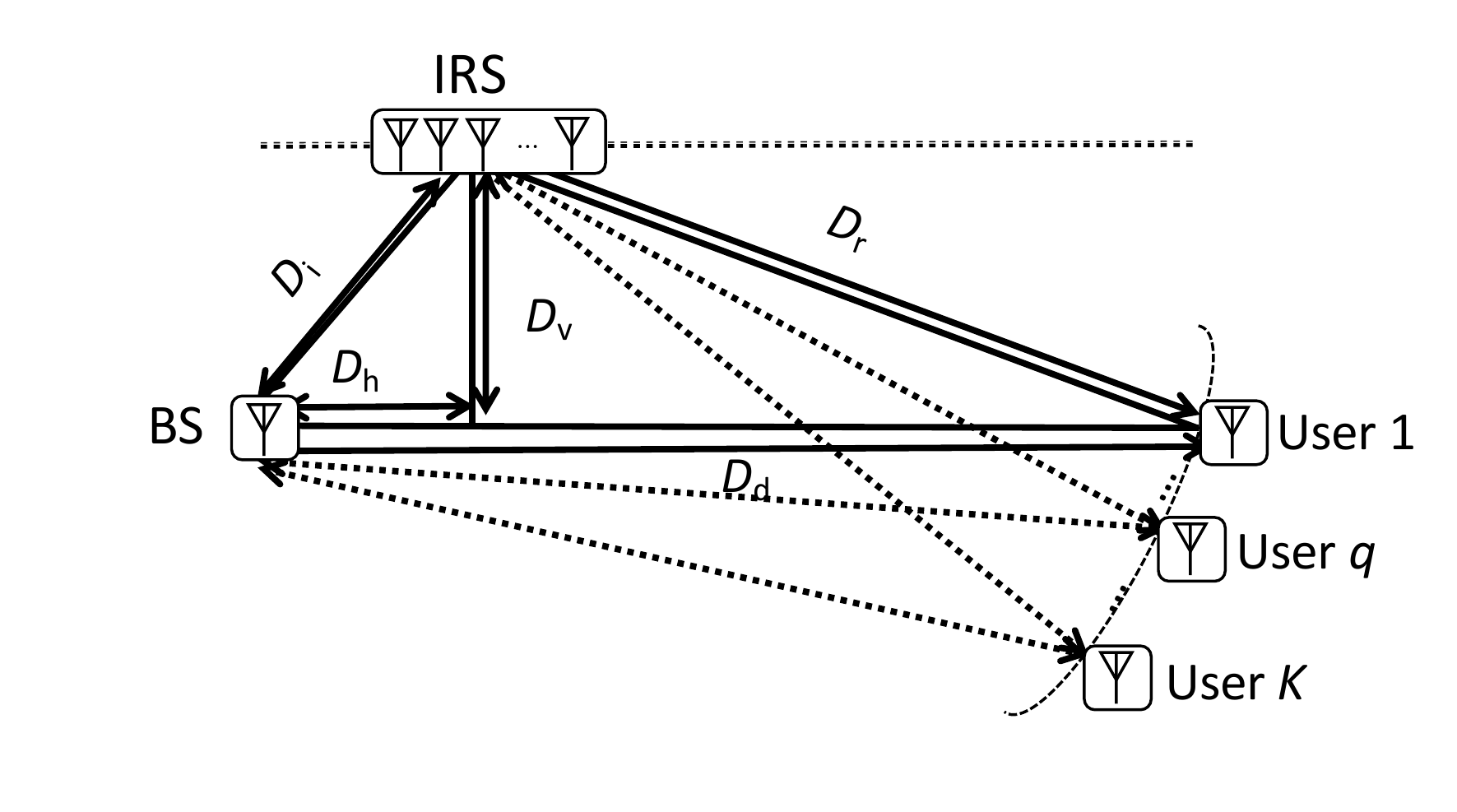}
    \vspace{-8mm}
    \caption{{Simulation layout.}}
   \label{fig: layout}
\end{figure}
\section{Numerical result}\label{Section_numerical_result}
{We now evaluate the performance in a typical open space WiFi-like environment under a
transmit power constraint of 36 dBm  at a central frequency of 5.18 GHz with signal bandwidth being fixed to 10MHz as a baseline unless otherwise stated. A uniform linear array (ULA) at the IRS with half wavelength spacing are considered in our model.  The pathloss model is shown below
\begin{equation}
    T_{j}(D_{j}) = r_0(\frac{D_j}{d_0})^{-\nu_j}
\end{equation}
where $r_0$ is the large scale fading parameter at reference distance $d_0 = 1$m with $D_j$ and $\nu_j$ referring to the distance and pathloss exponent, respectively, for $j\in\{\mathrm{d}, \mathrm{i}, \mathrm{r}\}$. All the channels are modeled as Rayleigh fading NLOS channels with
the path loss exponents and power 
delay profiles coming from  model D in \cite{erceg2004ieee}. 18 taps are modeled as i.i.d CSCG random variables 
to generate uncorrelated frequency-selective fading channels. }
According to Fig. \ref{fig: layout},  $D_\mathrm{i} = \sqrt{D_\mathrm{h}^2 + D_\mathrm{v}^2}$ and $D_\mathrm{r} = \sqrt{D_\mathrm{v}^2 + (D_\mathrm{d} - D_\mathrm{h})^2}$ with $D_\mathrm{h}$ and $D_\mathrm{v}$ referring   horizontal distance and vertical distance, respectively.
Parameters are assigned as $D_\mathrm{v} = 2$m, $D_\mathrm{h} = 2$m and $D_\mathrm{d} = 15$m as a baseline unless specified later.
Without loss of generality, the reference path loss for all users is set as -35dB at 1m.  For all the numerical figures, SU FS-IRS and MU FS-IRS simulation results come from  Algorithm \ref{alg:SU-FS-IRS} and \ref{alg:MU-FS-IRS}, respectively. Both SU FF-IRS and MU FF-IRS results are generated from  Algorithm \ref{alg:MU-FF-IRS} for  choices of $K = 1$ and $K > 1$, respectively. 
The sufficient small tolerance of stopping threshold in all algorithms is set as $10^{-4}$  and each point in the following figures is acquired via averaging over 1000 independent realizations. The number of candidates for Gaussian randomization in Algorithm \ref{alg:MU-FF-IRS} is 1000.

\subsection{Single-User}\label{section_com_sufffs}
We first characterize the average output DC current versus the number of sinewaves $N$ with different number of passive reflecting elements $L$  in Fig. \ref{fig: L11020Mt5} and \ref{fig: L151015202530Mt1N64}. A \textit{first}  observation is that the output DC current of both SU FS-IRS and SU FF-IRS  increases  with the  number of sinewaves $N$ and SU FS-IRS can be seen as a performance upper bound of the corresponding SU FF-IRS scenario. When  $N=1$,  SU FF-IRS displays exactly the same current with SU FS-IRS. A \textit{second} observation is that the output current approximately scales up with the  $L^4$. 
This is, thus, envisioned to revolutionize large-scale design since a large  $L$ can  compensate for the small number of transmit antennas. A \textit{third} observation is that the performance gap between SU FS-IRS and SU FF-IRS is gradually increased with  $N$  and SU FS-IRS  observes a gain of 28$\%$  over SU FF-IRS when $N = 64, L = 40$. The reason behind such a phenomenon is that, in SU FS-IRS, ideal passive beamforming phases can always align the  composite channel. While, in SU FF-IRS, more performance loss is incurred by the misalignment between the auxiliary channel and the direct channel with larger $N$.
\par An overview of average output DC current versus the number of sinewaves $N$ is clearly exhibited in Fig. \ref{fig: ASSN12481632M4L30}. SU FS-IRS, SU FF-IRS and SU Rand-Phase-IRS WPT refer to nonlinear-based rectenna model with different IRS scenarios. SU No-IRS WPT refers to conventional WPT designs in \cite{clerckx2016waveform}. SU FS-IRS ASS and SU No-IRS ASS denote the ASS strategy based linear-based rectenna model {(a truncating order of 2)} \cite{clerckx2016waveform}\cite{huang2017large} with FS-IRS and without IRS, respectively\footnote{{The output current for user $q$ under a linear based model (a truncating order of 2) is given as $i_{\textrm{dc},q}(\mathbf{s},\{\mathbf{\Theta}_n\}_{n=1}^{N})  =  \frac{1}{2}{k}_2\sum_{n=1}^{N}{s}_n^* {h}_{q,n}^* {h}_{q,n}{s}_n$. Compared with (\ref{for: intiobj}) , the maximization of output current can be efficiently solved by allocating all the power to the strongest frequency-selective channel which leads to the ASS strategy in Fig. \ref{fig: ASSN12481632M4L30}}.}. 
\begin{figure}[t]
    \centering
    \begin{minipage}[b]{0.45\textwidth} 
    \centering 
    \includegraphics[width = 7cm]{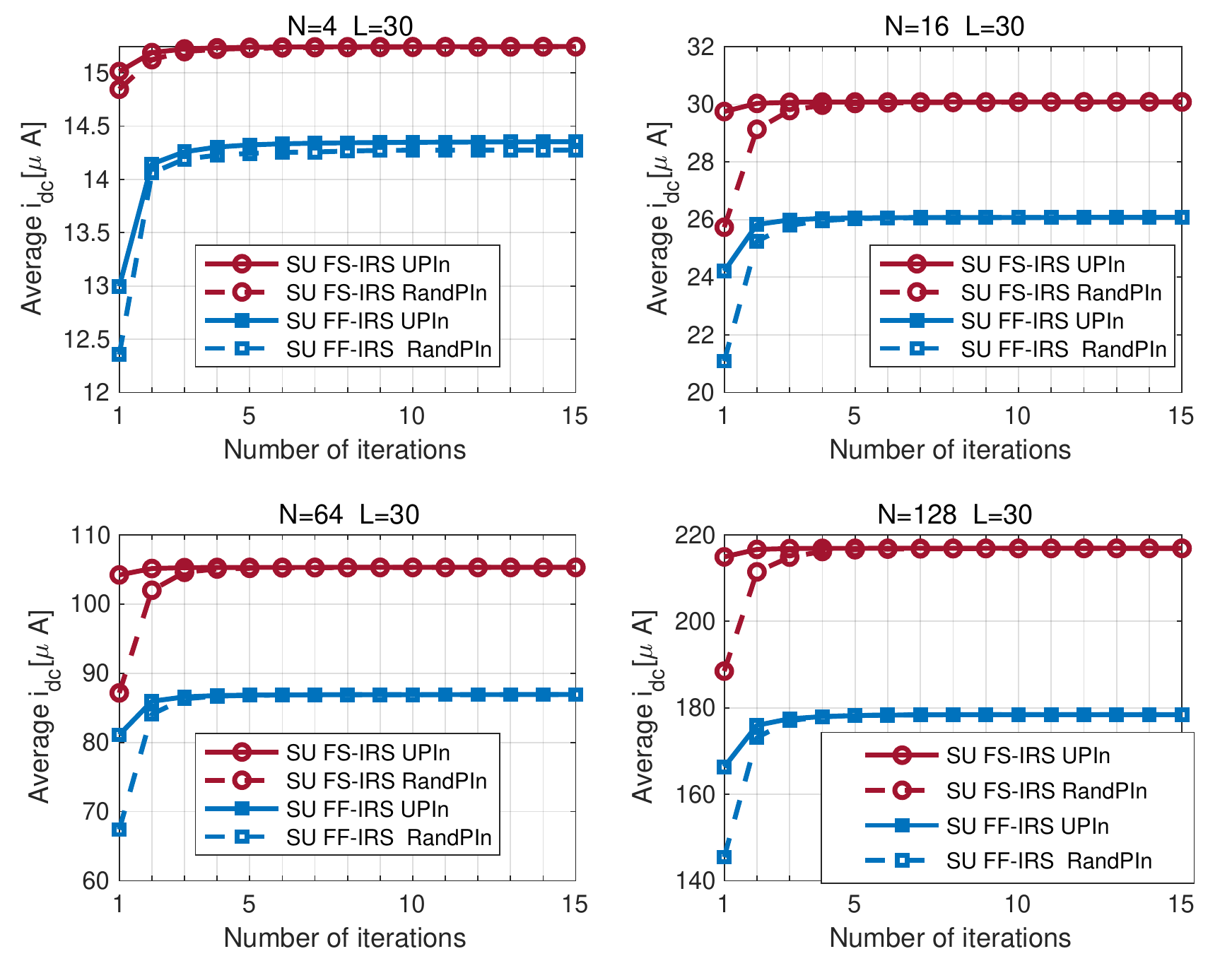}
    \caption{Average $i_{\textrm{dc}}$ versus number of iterations for  $N = 4,16,64,128$ and $L = 30$.}
   \label{fig: N1864256Mt4L10}
    \end{minipage} 
    \begin{minipage}[b]{0.45\textwidth}
    \centering
    \includegraphics[width = 7cm]{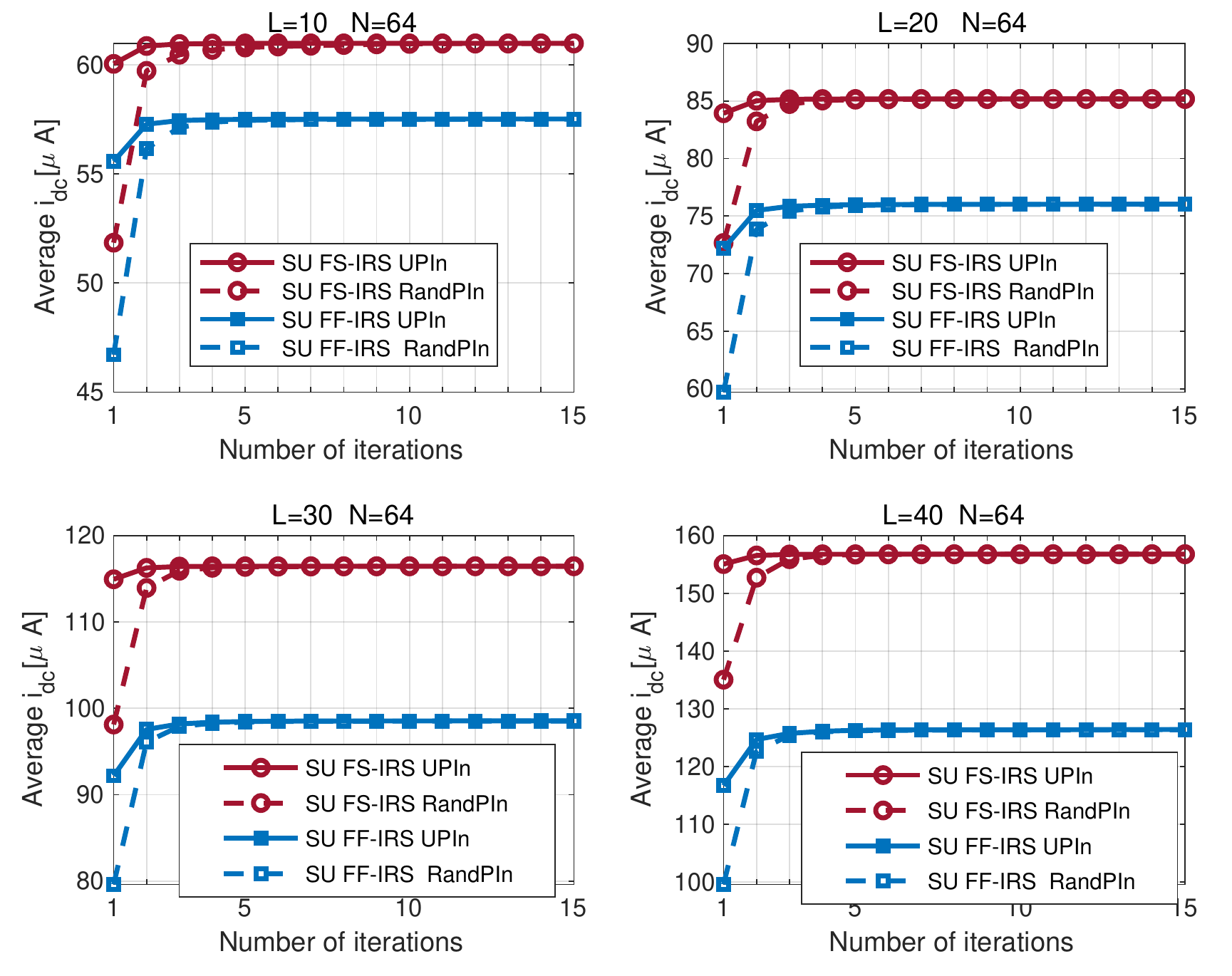}
    \caption{Average $i_{\textrm{dc}}$ versus number of iterations for  $L = 10,20,30,40$ and $N = 64$.}
   \label{fig: L10203040Mt4N64}
   \end{minipage}
\end{figure} 
We make the following observations. 
\par \textit{First}, if taking SU FS-IRS ASS WPT as a benchmark, SU FS-IRS and SU FF-IRS can exceed the benchmark on around 550$\%$ and  400$\%$, respectively, when $N = 64$.  This result strongly embodies the superiority of nonlinear based rectenna model over ASS based linear rectenna  model, showing that the output DC current can be effectively boosted by leveraging the gain from rectenna nonlinearity for both FS-IRS and FF-IRS scenarios. 
\par\textit{Second}, both SU FS-IRS and SU FF-IRS observe a great performance gain over No-IRS scenario  on around  195$\%$ and 155$\%$, respectively, when $N = 64, L = 30$. An explanation is that both of them consolidate the composite channel strength with passive beamforming phases manipulation. Furthermore, it is worth pointing out that although SU FF-IRS is outperformed by SU FS-IRS, it still achieves non-negligible performance improvement over No-IRS conventional WPT waveform designs in \cite{clerckx2016waveform}\cite{huang2017large}.
\par\textit{Third},  SU Rand-Phase-IRS displays a similar output current with No-IRS scenario which  emphasizes the significance of the passive beamforming design.  Without the optimization of the passive beamforming phases,  the signal attenuation  can not be effectively alleviated by IRS as different cascaded channels superpose randomly instead of constructively.


\subsection{Convergence Analysis}\label{subsection_convergence_analysis}
Fig. \ref{fig: N1864256Mt4L10} and \ref{fig: L10203040Mt4N64}
illustrate the average output DC current versus the number of iterations for different  initialization strategies. The initial passive beamforming phases are randomly chosen in $[0,2\pi)$. The frequency domain power allocation strategies 
are "UPIn" and "RandPIn"  referring  uniform power allocation and  random power allocation, respectively. First, "UPIn" exhibits a faster convergence 
than "RandPIn", which suggests the potential benefits of uniform power allocation in large scale SU FS-IRS and SU FF-IRS WPT. Second, both "UPIn" and "RandPIn" converge to  nearly the same  objective function values. An explanation is that considering the  non-convexity in problem (\ref{for: pff}), various initial points may lead to different  solutions in Algorithm \ref{alg:MU-FF-IRS}. However,  solution $\textbf{X}^{\star}$  is demonstrated to be a rank-1 solution in all tested channels and strongly reduces the performance loss incurred by Gaussian randomization method, which contributes to the insensitivity of initial power allocation and initial passive beamforming phases.

\subsection{Bandwidth}
\begin{figure}[t]
    \centering
    \begin{minipage}[b]{0.45\textwidth}
    \centering 
    \includegraphics[width = 8.2cm]{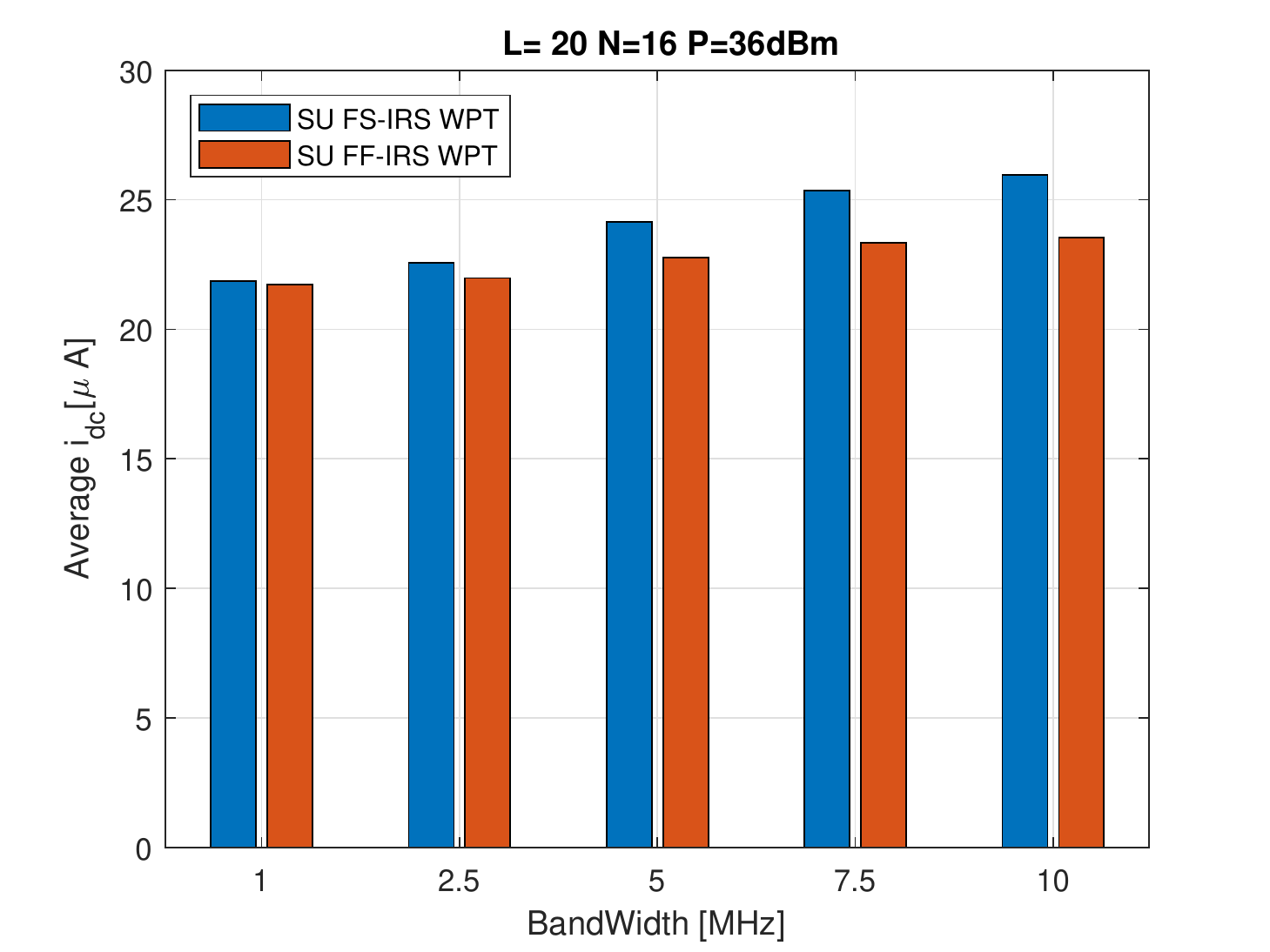}
    \caption{Average $i_{\textrm{dc}}$ versus   bandwidth for $L = 20$ and $N = 16$.}
   \label{fig: bandwidthBarMt2L20N32}
    \end{minipage}
    \begin{minipage}[b]{0.45\textwidth}
    \centering
    \includegraphics[width = 8.2cm]{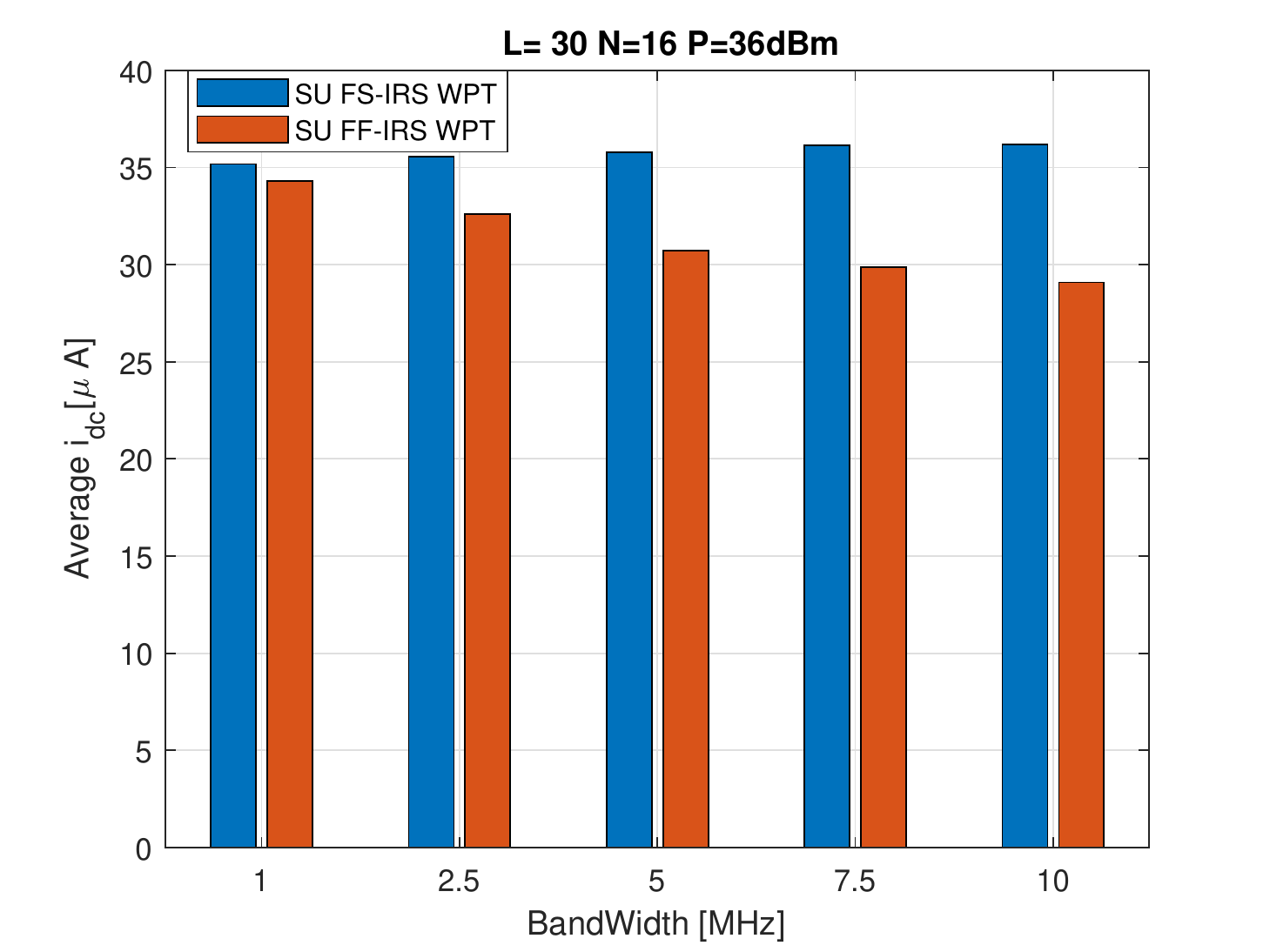}
    \caption{Average $i_{\textrm{dc}}$ versus   bandwidth for $L = 30$ and $N = 16$.}
   \label{fig: bandwidthBarMt2L30N32}
   \end{minipage}
\end{figure}
Fig. \ref{fig: bandwidthBarMt2L20N32} and Fig. \ref{fig: bandwidthBarMt2L30N32} explores the sensitivity of $i_{\textrm{dc}}$ to different bandwidths from  1MHz to 10 MHz for  $L = 20$ and  $L = 30$, respectively. A \textit{first} observation is that the gap between SU FS-IRS and SU FF-IRS in both figures increases  with the bandwidth which emphasizes the superiority of SU FS-IRS over SU FF-IRS on  output DC current by favouring the strongest sinewaves more efficiently. A \textit{second} observation is that the output DC current of SU FS-IRS increases slowly with the bandwidth and the opposite behavior is observed in SU FF-IRS. An explanation is that, when $L = 20$, both SU FS-IRS and SU FF-IRS benefit from increasing frequency selectivity which contributes to an ascending trend. However, when $L = 30$,  the misalignment of composite channel in SU FF-IRS becomes more pronounced with increasing bandwidth, which renders a descending trend. 
\par To conclude,  SU FF-IRS is not equally suitable for broadband and narrowband WPT, namely, SU FF-IRS works well in narrowband systems  but incurs some loss in a higher bandwidth transmission. In contrast, SU FS-IRS offers a promising  gain over SU FF-IRS by flexibly favouring passive beamforming phases in a frequency-selective design but also incurs a higher hardware complexity.

\subsection{Multi-User}
The average  weighted sum current  versus the number of sinewaves $N$  for different number of users $K$ is shown in Fig. \ref{fig: MUK135compare}. 
The weight $\xi_q$ for each user  is 1. It is first observed that MU FF-IRS incurs a performance loss compared with MU FS-IRS and the performance gap increases with $K$. This is expected since MU FF-IRS suffers more misalignment on different user's composite channels than MU FS-IRS by constraining the passive beamforming phases to be the same across frequencies, which  stresses the natural benefits of developing more flexible  IRS hardware to enable adaptive frequency-selective  passive beamforming phase shifts. 
\begin{figure}[t]
    \centering
    \begin{minipage}[b]{0.45\textwidth} 
    \centering 
    \includegraphics[width = 7cm]{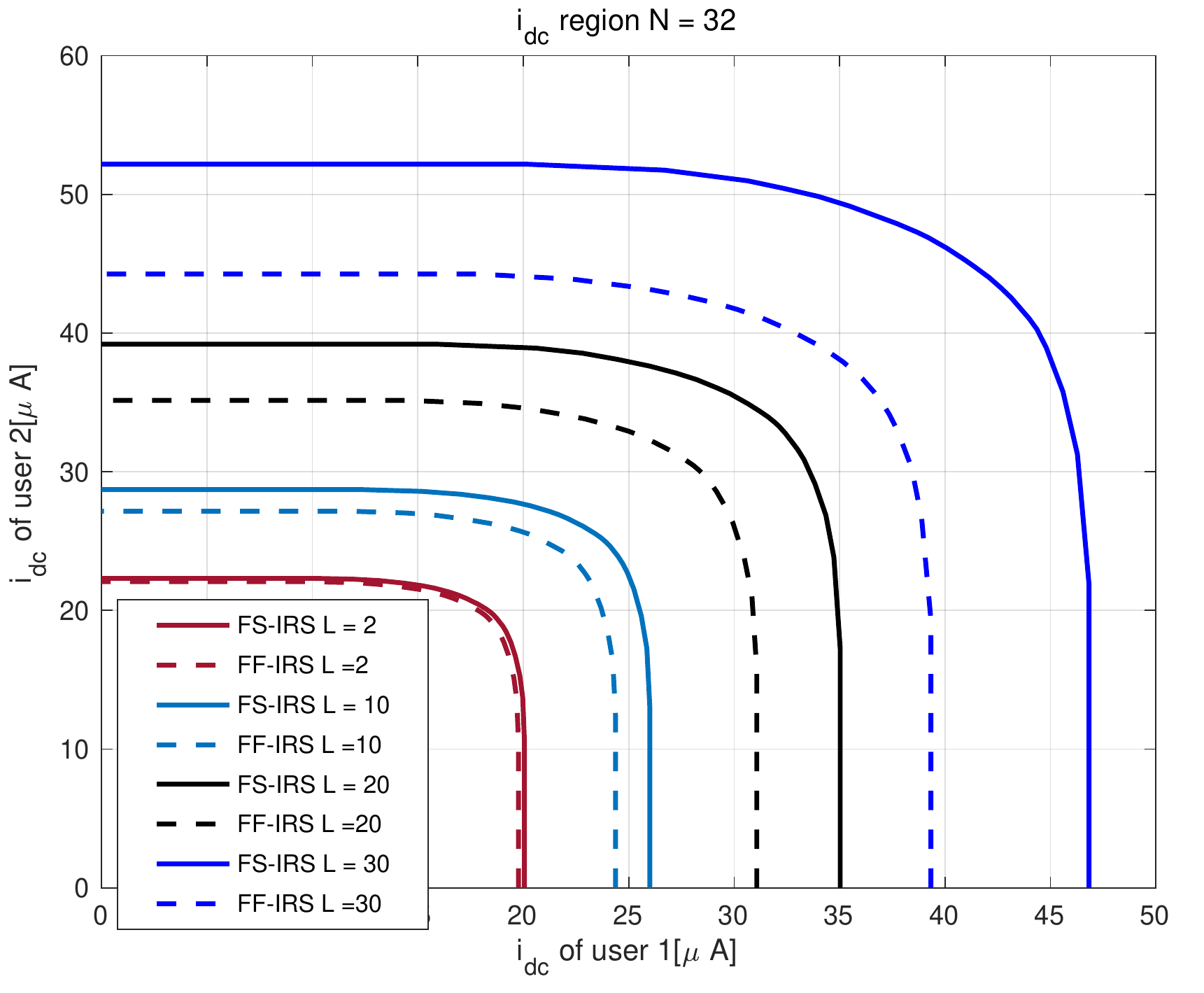}
    \caption{$i_{\textrm{dc}}$ region versus $L$ for $N = 32$.}
   \label{fig: rateRegionL2102030}
    \end{minipage} 
    \begin{minipage}[b]{0.45\textwidth}
    \centering
    \includegraphics[width = 7cm]{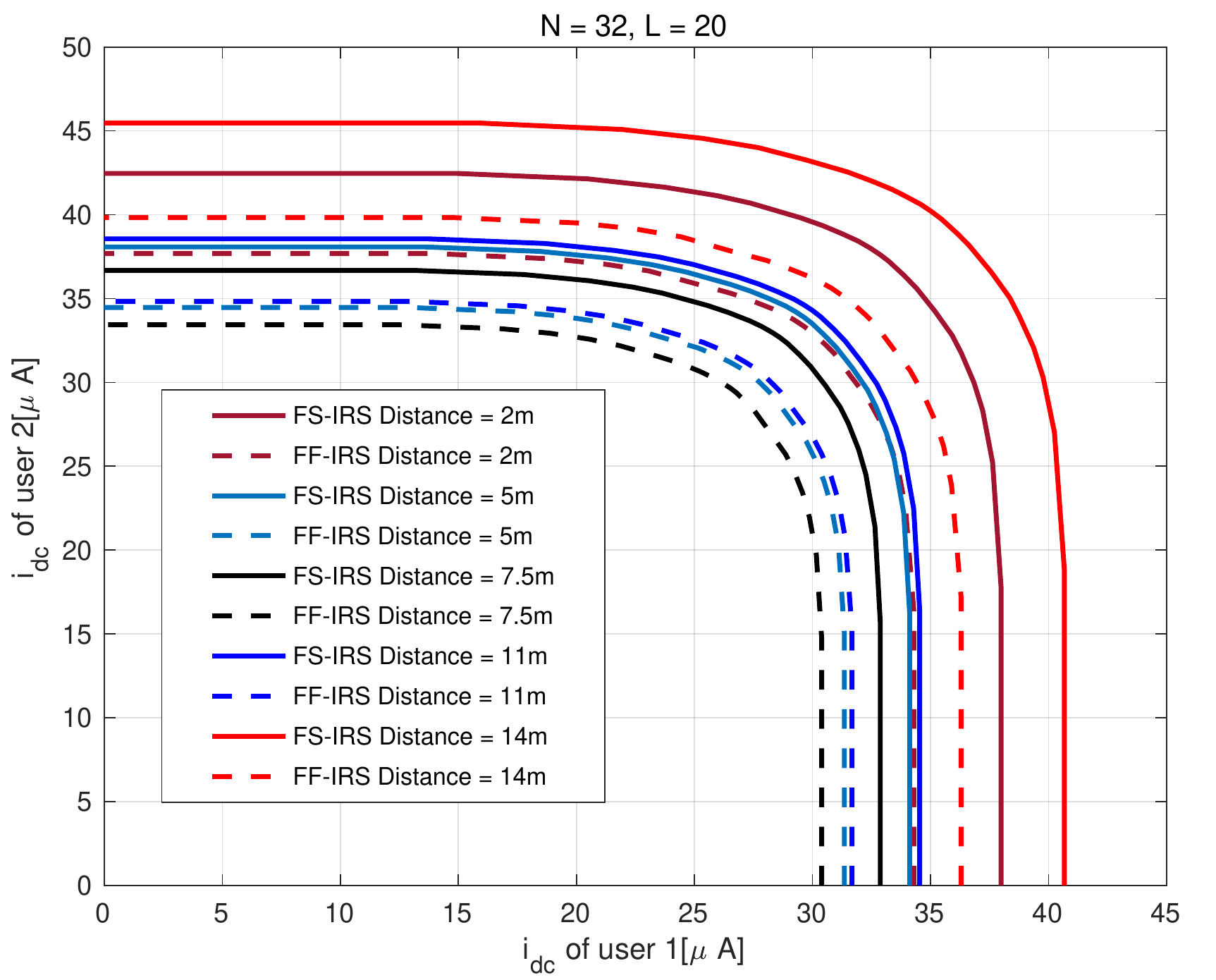}
    \caption{$i_{\textrm{dc}}$ region versus $D_\mathrm{h}$ for $N = 32$ and $L = 20$.}
   \label{fig: rateRegionDistance}
   \end{minipage}
\end{figure} 
\par
\begin{figure}[t]
    \centering
    \begin{minipage}[b]{0.45\textwidth}
    \centering 
    \includegraphics[width = 7.6cm]{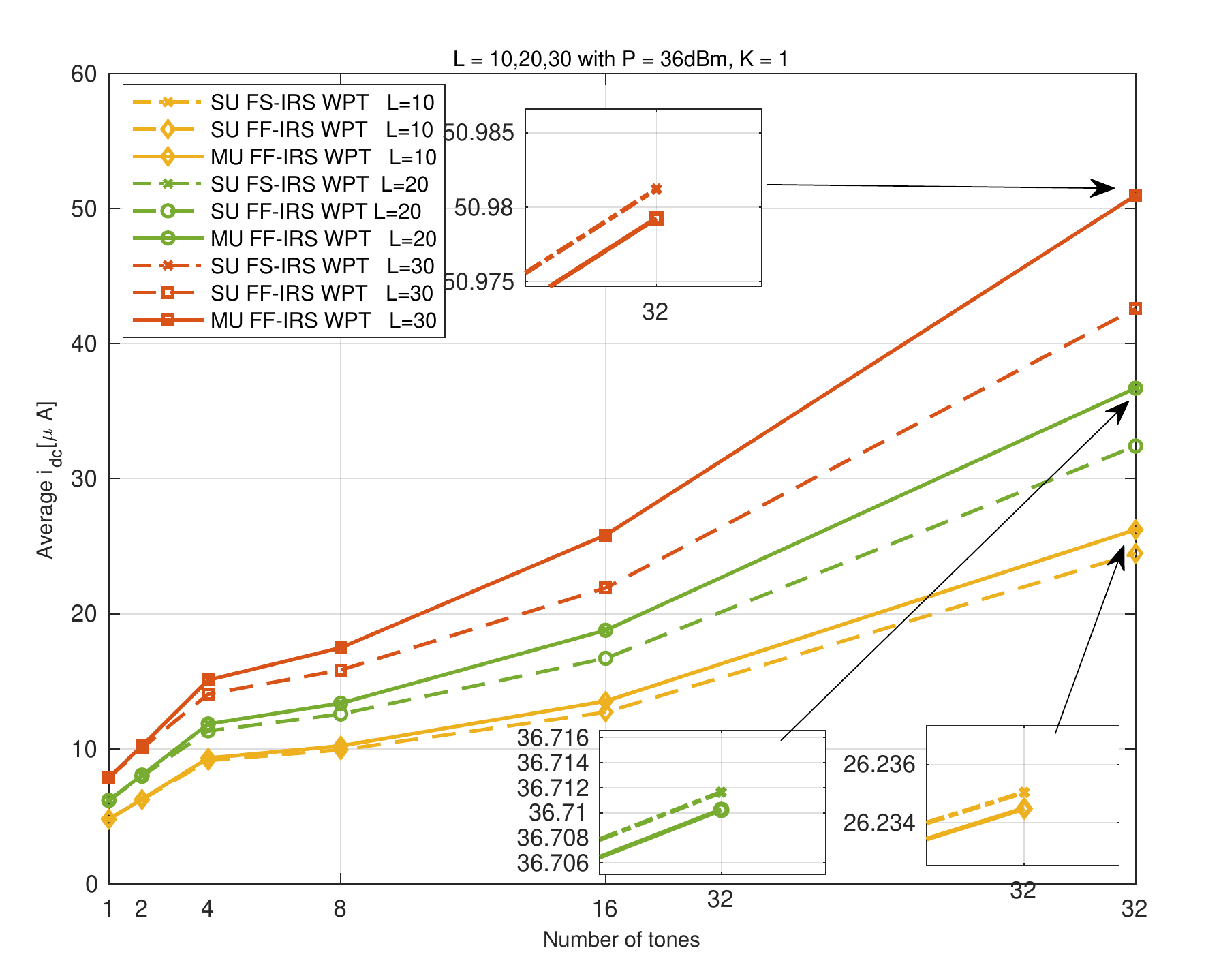}
    \caption{ {Average $i_{\textrm{dc}}$ versus    $N$ for $L = 10, 20, 30$ and $K = 1$.}}
   \label{fig: alg2alg3compareK1L102030P36N12481632}
    \end{minipage}
    \begin{minipage}[b]{0.45\textwidth}
    \centering
    \includegraphics[width = 7.6cm]{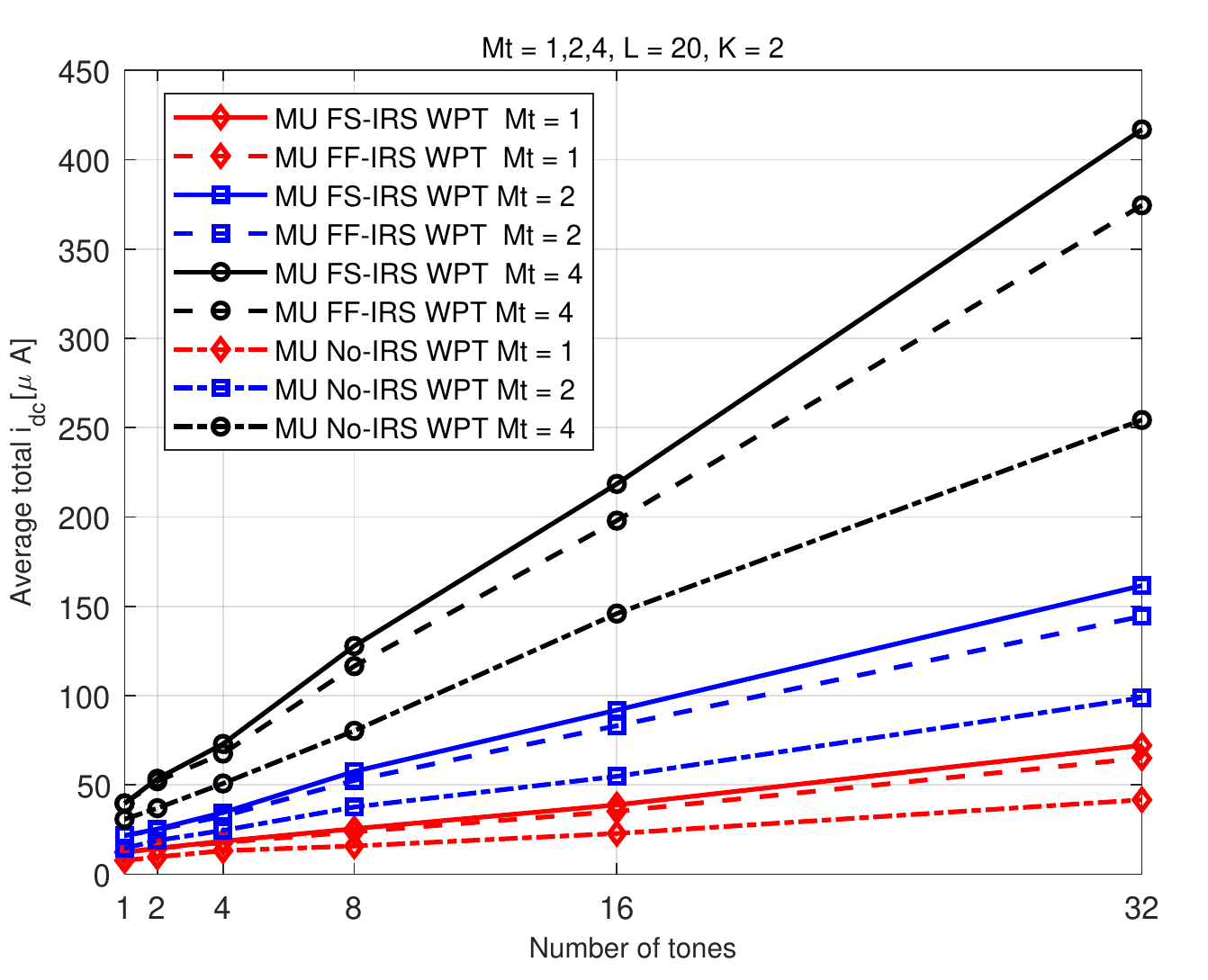}
    \caption{{Average weighted sum $i_{\textrm{dc}}$ versus   $N$ for $L = 20$, number of transmit antennas $M_t = 1, 2, 4$ and $K = 2$.}}
   \label{fig: MUK2L20Mt124P36N12481632}
   \end{minipage}
\end{figure}

Fig. \ref{fig: rateRegionL2102030} draws a useful insight into the average  $i_{\textrm{dc}}$ region (obtained by solving problem (\ref{for:InitialMUP}) with varying weights $(\xi_1,\xi_2)$) versus the number of reflecting elements $L$ in a two user system. It is observed that the current regions of two IRS strategies gradually increase with  $L$. This is intuitive since both  FS-IRS and  FF-IRS can compensate the distance-dependent path loss by strengthening  the composite channel of each user. Due to the additional design flexibility in the frequency domain, FS-IRS provides better channel alignment and it enlarges the region more than FF-IRS. 
This characteristic highlights the advancement of IRS in frequency-selective design. To further unveil the impact of the IRS position, based on the  same user weight pairs as Fig. \ref{fig: rateRegionL2102030}, Fig \ref{fig: rateRegionDistance} demonstrates the current region versus the horizontal distances, i.e. $D_\mathrm{h}$,  with $D_\mathrm{v} = 2$m  as a constant.  One can observe that the regions are effectively enlarged when IRS is installed either near the BS or near the users. {The comparison between Algorithm \ref{alg:MU-FS-IRS} and Algorithm \ref{alg:SU-FS-IRS} with $K = 1$ is demonstrated in Fig. \ref{fig: alg2alg3compareK1L102030P36N12481632}. Due to suboptimal operation of EWU strategy, Algorithm \ref{alg:MU-FS-IRS} exhibited a slightly  loss in contrast with Algorithm \ref{alg:SU-FS-IRS}. However, this loss is small enough to be negligible (less than $0.1\%$) which, instead, reveals the generality and superiority of Algorithm \ref{alg:MU-FS-IRS}.}
{The impact of joint waveform, active and passive beamforming is demonstrated in Fig. \ref{fig: MUK2L20Mt124P36N12481632} which explores the average weighted sum current versus the number of sinewaves $N$ for different number of transmit antenna $M_t$. It is observed that the weighted sum output current of both MU FF-IRS and MU FS-IRS approximately scale up with the square number of $M_t$. Compared with single-antenna scenario in Fig. \ref{fig: L151015202530Mt1N64}, active beamforming is envisioned to further boost the output DC power.}
\subsection{{Discrete-Phase}}
\begin{figure}[t]
    \centering
    \begin{minipage}[b]{0.45\textwidth}
    \centering 
    \includegraphics[width = 7.5cm]{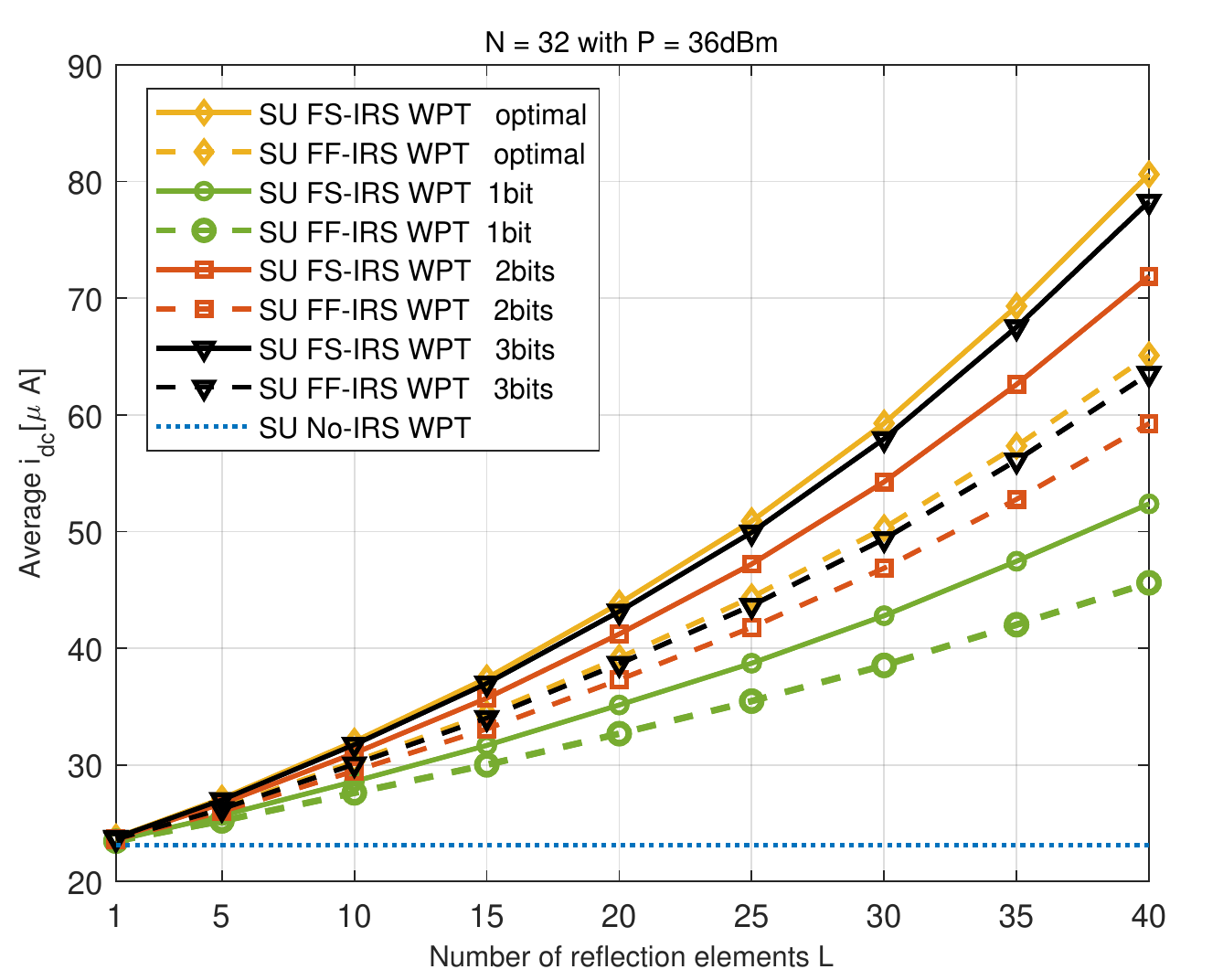}
    \caption{{Average  $i_{\textrm{dc}}$ versus   $L$ for $N = 32$ and resolution bits $M = 1, 2, 3$.}}
   \label{fig: SUdiscrete123L1to40N32P36}
    \end{minipage}
    \begin{minipage}[b]{0.45\textwidth}
    \centering
       \includegraphics[width = 7.8cm]{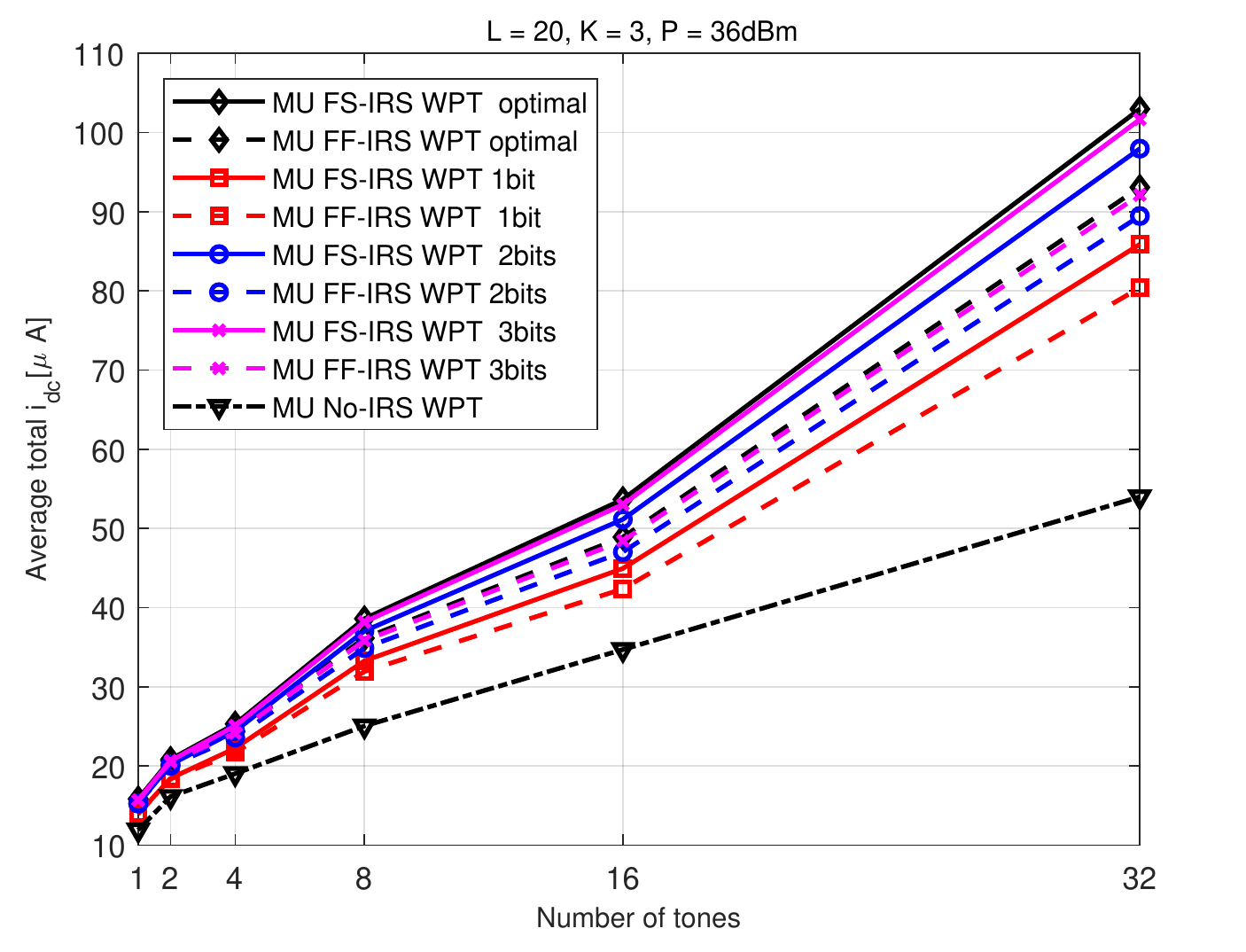}
    \caption{ {Average weighted sum $i_{\textrm{dc}}$ versus    $N$ for $M = 1, 2, 3$, $L = 20$ and $K = 3$.}}
   \label{fig: MUK3discrete123L20N12481632}
   \end{minipage}
\end{figure}

{To extend to the discrete-phase IRS problem,  one widely used strategy is to relax the discrete IRS problems to their continuous counterparts and obtain quantized phases $\tilde{\psi}_{l + (n-1)L} \forall n, l$ in phase set $\mathcal{S}$ by mapping the continuous values to their closest  discrete value in  $\mathcal{S}$ with mapping function $\mathcal{F}$ \cite{wu2019beamforming} which  is given as
\begin{equation}
    \tilde{\psi}_{l + (n-1)L} = \mathcal{F}(\psi_{l + (n-1)L}), \forall l, n, \label{map}
\end{equation}
\begin{equation}
   \tilde{ \psi}_{l + (n-1)L} \in \mathcal{S}, \forall l,n,
\end{equation}
where $\mathcal{S} = \{0,  \Delta \psi, \dots, \Delta \psi (2^{M} - 1)\}$ denotes the equally spaced phase set with  $\Delta \psi = 2\pi/2^{M}$ and $M$ denotes the resolution bits. Leveraging this discrete-phase strategy in \cite{wu2019beamforming}, Fig. \ref{fig: SUdiscrete123L1to40N32P36} and \ref{fig: MUK3discrete123L20N12481632} are plotted for single-user and multi-user conditions, respectively.}

{Fig. \ref{fig: SUdiscrete123L1to40N32P36}  demonstrates the average output DC current versus  $L$ with $N = 32$ and quantization scheme $M = 1, 2, 3$ bits for single-user scenario.  \textit{First}, it is observed that the $i_{\mathrm{dc}}$ for both SU FF-IRS and SU FS-IRS can be greatly improved  compared with No-IRS condition even with 1-bit phase shifter. 
\textit{Second}, one can observe that the performance loss incurred by discrete phases decreases with an increasing number of resolution bits. This is intuitive since a larger number of  bits enables a better alignment between channels. Under the same quantization scheme, right figure illustrates the weighted sum current versus $N$ with $L = 20$ and $K = 3$. The performance gain of 1-bit resolution current over No-IRS benchmark increases with the number of subcarriers $N$, which suggests the effectiveness of deploying discrete-phase IRS in wideband WPT. Moreover, the performance loss incurred by  discrete-IRS  is low with 3-bit resolution phase shifters, which confirms that directly quantizing the optimized continuous RE phases  can achieve near-optimal performance in multi-user scenario.  }

\section{{Conclusion and Future Work}}\label{Section_conclusion}
In this paper, we investigated the joint waveform and passive beamforming design for IRS-aided wireless power transfer for both single-user and multi-user deployments. Two different formulations were developed based on a FS-IRS and FF-IRS. An optimization framework based on AO and SCA was studied and demonstrated  robustness with fast convergence. Numerical results confirmed the inefficiency of linear based model and highlight the fact that even FF-IRS-aided WPT can outperform conventional WPT  designs with IRS passive beamforming gain. Furthermore, FS-IRS acted as a canonical upper bound for broadband system while FF-IRS was suitable for narrowband transmission. Moreover, for WPT system, the assistance of IRS  not only expanded the operation range  for SU conditions, but also enlarged the output DC current region in MU scenarios. {In addition, this AO framework can be directly extended to multi-antenna scenario to explore the transmit beamforming gain. Last but not least, a near-optimal result can be achieved by leveraging low-resolution discrete-phase IRS for both SU and MU scenarios}.

{Several  important issues are not addressed in our paper: first, how to formulate the waveform design and passive beamforming for MU minimum $i_{\textrm{dc}}$ maximization problem (Max-Min problem) \cite{huang2017large}; second, how to involve more advanced  group/fully connected  IRS into WPT system \cite{shen2020modeling}; third, how to design waveform, transmit active beamforming, receive
  combining and passive beamforming in a multi-user MIMO WPT scenario \cite{shen2020joint}; fourth, how to relax the constraint on assumptions of the EH model (ideal low pass filter with infinite RC constant) and explore this more accurate and complicated rectenna model by using machine learning techniques \cite{clerckx2021wireless}.} 

\appendices
\section{{Proof of Proposition \ref{Proposition_alg2_phase}}}\label{appendixA}
{We first prove the convergence of  problem (\ref{pro: 36p}).
${f}(\mathbf{d}_q)$ and ${f}(\mathbf{d}_q,\mathbf{d}_q^{(i-1)})$ are demonstrated below
\begin{equation}
    {f}(\mathbf{d}_q) =  \mathbf{d}_q^{H}\mathbf{K}_0\mathbf{d}_q,\label{for:SCAaffine3}
\end{equation}
\begin{equation}
    {f}(\mathbf{d}_q,\mathbf{d}_q^{(i-1)}) = 2\Re\{\mathbf{d}_q^{(i-1)H}\mathbf{K}_0\mathbf{d}_q\}-\mathbf{d}_q^{(i-1)H}\mathbf{K}_0\mathbf{d}_q^{(i-1)}.\label{for:SCAaffine4}
\end{equation} 
According to  \cite{mehanna2014feasible}, it can be checked that $ f(\mathbf{d}_q^{(i)},\mathbf{d}_q^{(i)}) =  f(\mathbf{d}_q^{(i)})$.
Since the solution vector $\mathbf{d}_q^{(i - 1)}$ is a feasible solution in iteration $i-1$, by using the inequality in  Taylor expansion, we have $f(\mathbf{d}_q^{(i)}) \geq f(\mathbf{d}_q^{(i)},\mathbf{d}_q^{(i-1)})$ which proves that $ f(\mathbf{d}_q^{(i)})$ is monotonically increasing. Additionally, problem (\ref{pro: 36p}) must have a upper bound due to the unit modulus constraint (\ref{for: p16b}). Hence, problem (\ref{pro: 36p}) is guaranteed to converge.}
    
{Now, we prove that the convergent solution is a KKT solution of problem (\ref{for: pff}). 
(\ref{for: p16a}) and (\ref{for: p16b}) are convex problems with respect to  $\mathbf{X}$ and satisfies the slater’s condition \cite{boyd2004convex}, the dual gap is zero and strong duality hold. The optimal solution  $\mathbf{X}^{\star}$ can be get by figuring out its dual problem. The corresponding Lagrange function of problem (\ref{pro: 36p}) is 
\begin{equation}
    \mathcal{L}(\mathbf{X}, \bm{\tau}) = \text{Tr}\{\mathbf{K}_1\mathbf{X}\}  + \sum_{l^{\prime} = 1}^{L + 1}\tau_{l^{\prime}}( \mathbf{X}_{l^{\prime},l^{\prime}} - 1) + \mathbf{\Upsilon}\mathbf{X}
\end{equation}
where $\bm{\tau} = \{\tau_1, \tau_2, \dots, \tau_{L + 1}\}$ and $\mathbf{\Upsilon}$ denote the vector and matrix dual variables of  (\ref{for: p16b}) and (\ref{for: p16c}). There must be a  $\bm{\tau}^{\star} = \{\tau_1^{\star}, \tau_2^{\star}, \dots, \tau_{L + 1}^{\star}\}$ and $\mathbf{\Upsilon}^{\star}$ for guaranteeing that the  corresponding KKT conditions 
are satisfied
\begin{equation}
   \nabla_{\mathbf{X}^{\star}}\mathcal{L}(\mathbf{X}, \bm{\tau})\vert_{\mathbf{X} = \mathbf{X}^{\star}} = \mathbf{0}\label{kkt1}
\end{equation}
\begin{equation}
    \tau_{l^{\prime}}^{\star}( \mathbf{X}^{\star}_{l^{\prime},l^{\prime}} - 1) = 0, \forall l^{\prime}\label{kkt2}
 \end{equation}
 \begin{equation}
    \mathbf{\Upsilon}^{\star}\mathbf{X}^{\star} = \mathbf{0}\label{kkt3}
 \end{equation}
The KKT conditions of problem (\ref{pro: 36p}) exactly consist of (\ref{kkt1}), (\ref{kkt2}) and (\ref{kkt3}). 
Hence, \textit{Proposition} \ref{Proposition_alg2_phase} holds.}

\section{{Proof of Proposition \ref{Proposition_alg1_phase}}}\label{appendixB}
{Due to the equivalence between problem (\ref{pro: sdr16}) and problem (\ref{pro: 36p}), (\ref{kkt1})-(\ref{kkt3}) are also the KKT conditions of problem (15).
From (\ref{for:SCAaffine2}), we have
\begin{equation}
      \nabla_{\mathbf{d}_q = \mathbf{d}_q^{\star}}{f}(\mathbf{d}_q)\vert_{\mathbf{d}_q = \mathbf{d}_q^{\star}} = \nabla_{\mathbf{d}_q = \mathbf{d}_q^{\star}}{f}(\mathbf{d}_q,\mathbf{d}_q^{(i-1)})\vert_{\mathbf{d}_q = \mathbf{d}_q^{\star}}\label{kkt4}
\end{equation}
Since problem (\ref{for: pffnonconv}) and problem (\ref{for: ffirsconv})  exactly contain ${f}(\mathbf{d}_q)$ and ${f}(\mathbf{d}_q,\mathbf{d}_q^{(i-1)})$ at iteration $i$, respectively,  except some constants, upon denoting the (\ref{for: pffnonconv}) as $p(\mathbf{d}_q) = \sum_{q=1}^{K}\xi_q(-\frac{1}{2}k_2{d}_{q,0}-{f}(\mathbf{d}_q))$ and (\ref{for: ffirsconv}) as ${p}(\mathbf{d}_q,\mathbf{d}_q^{(i-1)}) = \sum_{q=1}^{K}\xi_q(-\frac{1}{2}k_2d_{q,0}- f(\mathbf{d}_q,\mathbf{d}_q^{(i-1)}))$, we arrive at  
\begin{equation}
    \nabla_{\mathbf{d}_q = \mathbf{d}_q^{\star}}{p}(\mathbf{d}_q)\vert_{\mathbf{d}_q = \mathbf{d}_q^{\star}} = \nabla_{\mathbf{d}_q = \mathbf{d}_q^{\star}}{p}(\mathbf{d}_q,\mathbf{d}_q^{(i-1)})\vert_{\mathbf{d}_q = \mathbf{d}_q^{\star}}.
\end{equation}
According to (\ref{kkt4}) and relation between $p(\mathbf{d}_q)$ and ${p}(\mathbf{d}_q,\mathbf{d}_q^{(i-1)})$,
KKT conditions of problem (\ref{for: pff}) exactly consist of (\ref{kkt1}), (\ref{kkt2}) and (\ref{kkt3}) and $\mathbf{X}^{\star}$ is also a local optimal solution for problem (\ref{for: pff}). Combining the proof above and  \textit{Proposition}  \ref{Proposition_alg23_waveform},   a local optimal solution for passive beamforming subproblem and a global optimal solution for waveform design subproblem can be obtained. Hence, the AO Algorithm \ref{alg:MU-FF-IRS} converges to a local optimal solution in original problem (\ref{for:InitialMUP}) and
 \textit{Proposition}  
\ref{Proposition_alg1_phase} holds.}

\bibliographystyle{IEEEtran} 
\bibliography{main.bib}

\end{document}